\documentclass[11pt,a4paper,twoside]{article}
\pdfoutput=1
\usepackage[latin1]{inputenc}
\usepackage[pdftex]{graphicx}
\usepackage{subfigure, bbm}
\usepackage{amsmath,amsfonts,amssymb}
\usepackage{mathrsfs}
\usepackage[footnotesize]{caption}

\allowdisplaybreaks[1]

\begin{document}

\begin{titlepage}

\vspace*{-15mm}
\begin{flushright}
MPP-2008-31\\
\end{flushright}
\vspace*{0.7cm}

\begin{center}
{
\bf\LARGE
Quark and lepton masses at the GUT scale\\[2mm] 
including SUSY threshold corrections
}
\\[8mm]
S.~Antusch$^{\star}$
\footnote{E-mail: \texttt{antusch@mppmu.mpg.de}}, 
M.~Spinrath$^{\star}$
\footnote{E-mail: \texttt{spinrath@mppmu.mpg.de}},
\\[1mm]

\end{center}
\vspace*{0.50cm}
\centerline{$^{\star}$ \it 
Max-Planck-Institut f\"ur Physik (Werner-Heisenberg-Institut),}
\centerline{\it 
F\"ohringer Ring 6, D-80805 M\"unchen, Germany}
\vspace*{1.20cm}

\begin{abstract}

\noindent 
We investigate the effect of supersymmetric (SUSY) threshold corrections 
on the values of the running quark and charged lepton masses at the GUT 
scale within the large $\tan\beta$ regime of the MSSM. In addition to the 
typically dominant SUSY QCD contributions for the quarks, we also include 
the electroweak contributions for quarks and leptons and show that they 
can have significant effects. We provide the GUT scale ranges of quark 
and charged lepton Yukawa couplings as well as of the ratios $m_\mu/m_s$, 
$m_e/m_d$, $y_\tau/y_b$ and $y_t/y_b$ for three example ranges of SUSY 
parameters. We discuss how the enlarged ranges due to threshold effects 
might open up new possibilities for constructing GUT models of fermion 
masses and mixings.
\end{abstract}

\end{titlepage}

\newpage
\setcounter{footnote}{0}

\section{Introduction}

The unification of the fundamental forces of the Standard Model (SM) is one of the guiding principles in the search for a more fundamental theory of nature. In addition to providing a unified origin of the gauge interactions, Grand Unified Theories (GUTs), based e.g.\ on the gauge symmetry groups SU(5) \cite{Georgi:1974sy} or SO(10) \cite{Georgi:1974my}, also unify quarks and leptons of the SM in representations of the unified gauge groups. This property makes them attractive frameworks to address the flavour problem, i.e.\ the question of the origin of the observed pattern of fermion masses and mixings. Another attractive feature of left-right symmetric GUTs is the appearance of right-handed neutrinos in their particle spectra, which become massive after spontaneous symmetry breaking to the SM and thereby lead to the small observed neutrino masses via the seesaw mechanism \cite{seesaw}. In order to solve the gauge hierarchy problem inherent in high-energy extensions of the SM and to make the running gauge couplings meet (at the so-called GUT scale $M_{\mathrm{GUT}} \approx 2 \times 10^{16}$ GeV), the idea of Grand Unification is typically combined with that of low-energy supersymmetry (SUSY).

In order to construct successful GUT models of flavour, it is desirable to know the approximate GUT scale values of the quark and lepton masses and mixing parameters. The experimental data on the masses of the strange quark and the muon, for example, extrapolated to the GUT scale by means of renormalisation group (RG) running within the SM, gave rise to the so-called Georgi Jarlskog (GJ) relations \cite{Georgi:1979df} of $m_\mu/m_s = 3$ and $m_e/m_d = 1/3$ at the GUT scale, which can be realised from a Clebsch-Gordan factor after GUT symmetry breaking. The GJ relations have become a popular building block in many classes of unified flavour models.
In SUSY GUTs, another intriguing possibility emerges, which is the unification of all third family Yukawa couplings, i.e.\ of $y_t$, $y_b$, $y_\tau$ and furthermore, in the context of the seesaw mechanism, of $y_\nu$ at the GUT scale. To investigate whether this relation can be realised in a given model of low-energy SUSY, it is well known that a careful inclusion of SUSY threshold corrections is required \cite{Hall:1993gn, Carena:1994bv, Hempfling:1993kv, Blazek:1995nv}. These threshold effects are particularly relevant in the case of large $\tan \beta$, where $y_t = y_b = y_\tau$ seems achievable. 
Despite the possible importance of the SUSY threshold effects, in studies which interpolate the running fermion masses to the GUT scale these effects are typically ignored (see, e.g.,~\cite{Fusaoka:1998vc,Xing:2007fb}).  

The effects of SUSY threshold corrections on the possibility of third family Yukawa unification $y_t = y_b = y_\tau$ (and also on the less restrictive possibility $y_b = y_\tau$ which often emerges in SU(5) GUTs) has been extensively studied in the literature (see, e.g.,~\cite{Carena:1994bv, Hempfling:1993kv, Bagger:1996ei,King:2000vp, Blazek:2002ta}). Furthermore, recent studies \cite{Altmannshofer:2008vr} have addressed the phenomenological viability of this relation and have pointed out that under certain assumptions on the soft breaking parameters at the  GUT scale, $y_t = y_b = y_\tau$ may be already quite challenged by the experimental data from B physics. 
SUSY threshold effects on the GJ relations have been discussed recently in \cite{Ross:2007az}. Taking into account the typically leading contribution from SUSY QCD loops, it has been shown that the GJ relation can be realised under certain conditions on the SUSY parameters which govern this correction. 

The main purpose of this study is to include SUSY threshold corrections and calculate the possible GUT scale ranges for quark and charged lepton Yukawa couplings as well as for the ratios $m_\mu/m_s$, $m_e/m_d$, $y_\tau/y_b$ and $y_t/y_b$, which are important input parameters for GUT model building. 
Regarding $m_\mu/m_s$, compared to \cite{Ross:2007az} we include additional corrections from electroweak loops with binos and winos for quarks and charged leptons, which, as we will show, can have significant impact. Furthermore, instead of trying to fit the GJ relations by a sparticle spectrum, our aim is to analyse which alternative GUT scale relations may be possible and whether the GJ relation lies within the projected GUT scale ranges. We also investigate other relevant quantities of interest for GUT model building, such as, for example, the ratio of electron mass over down quark mass, $m_e/m_d$, and of the third family Yukawa couplings $y_t,y_b$ and $y_\tau$. 

The remainder of the paper is organised as follows: In section 2 we discuss the various $\tan\beta$-enhanced contributions to the SUSY threshold corrections and how we implement them within the renormalisation group running procedure of the Yukawa couplings. Section 3 contains our results for the GUT scale ranges for the  quark and charged lepton Yukawa couplings as well as for the ratios $m_\mu/m_s$, $m_e/m_d$, $y_\tau/y_b$ and $y_t/y_b$, and in section 4 we discuss possible implications for GUT model building. Section 5 concludes the paper.

\section{SUSY Threshold Corrections}

\begin{figure}[t]
 \centering
 \includegraphics[scale=0.65]{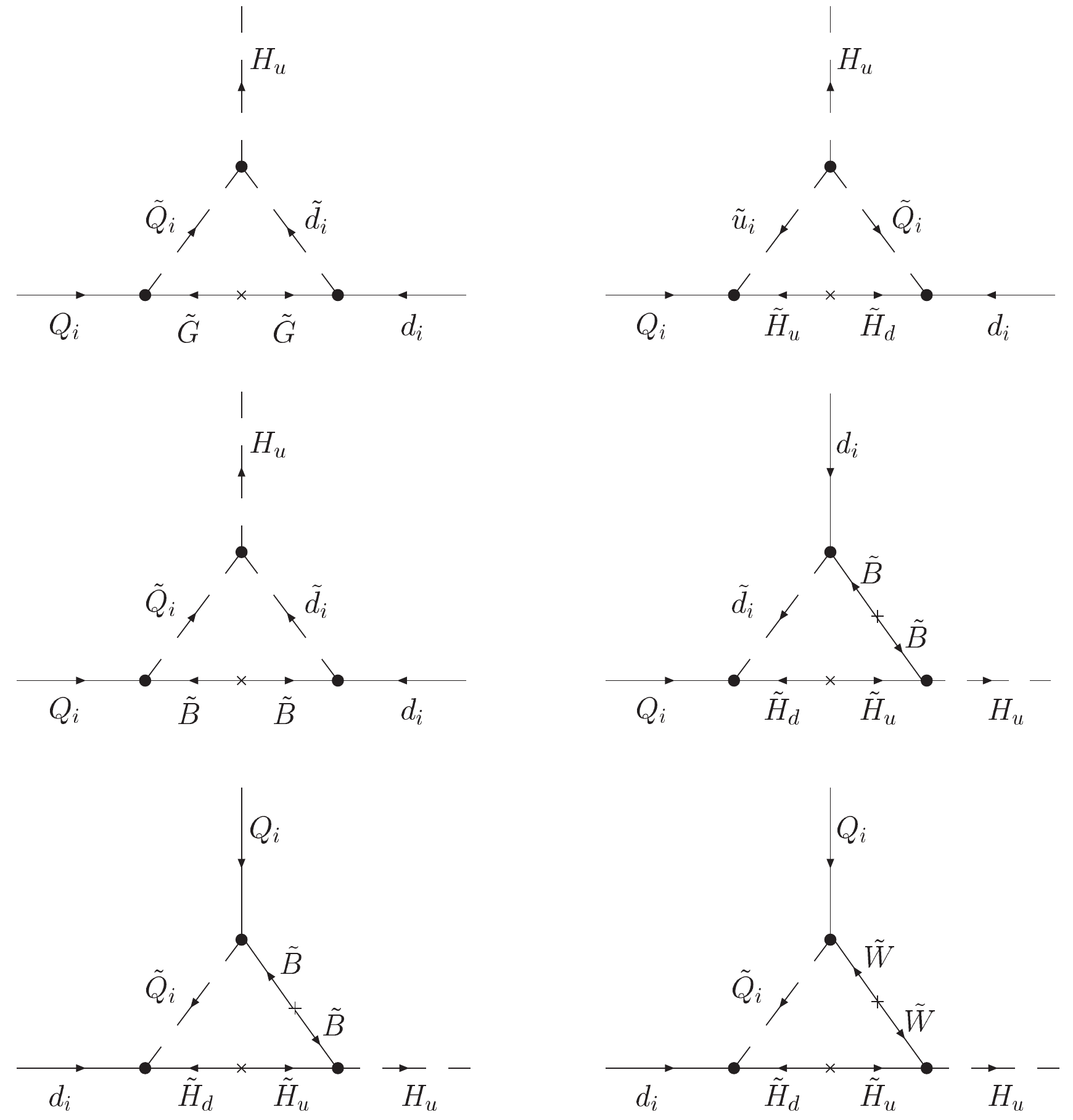}
 \caption{Feynman diagrams contributing to the SUSY threshold corrections to down-type quark Yukawa couplings. \label{fig:quarks}}
\end{figure}

\begin{figure}
 \centering
 \includegraphics[scale=0.65]{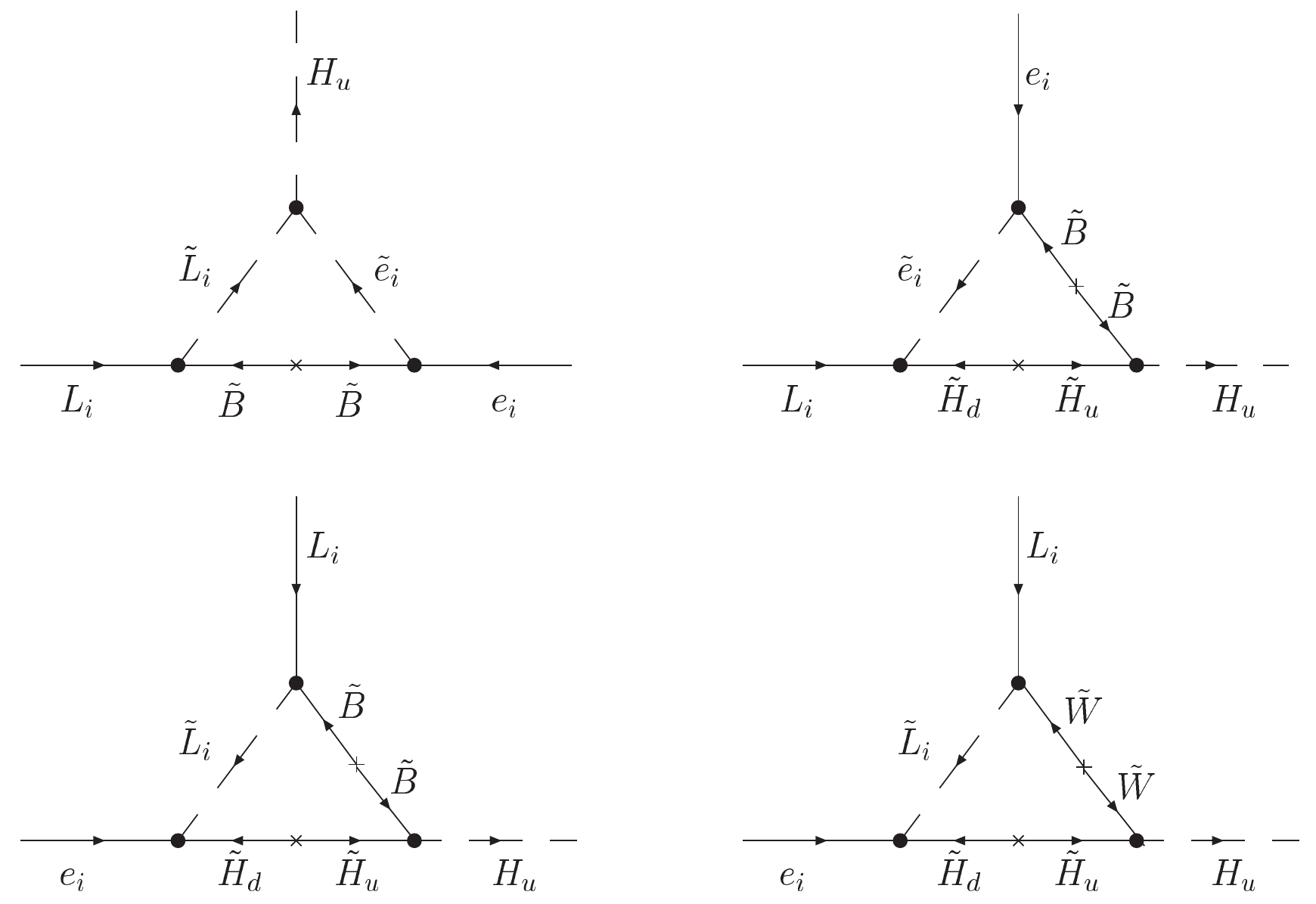}
 \caption{Feynman diagrams contributing to the SUSY threshold corrections to charged lepton Yukawa couplings. \label{fig:leptons}}
\end{figure}

For large $\tan \beta$ in the MSSM it is well known that certain supersymmetric one-loop corrections are enhanced (see, e.g.,\ \cite{Hall:1993gn, Carena:1994bv, Hempfling:1993kv}). The Feynman diagrams corresponding to these one-loop corrections are shown in figures \ref{fig:quarks} and \ref{fig:leptons}. $Q_i$ denotes the quark doublets, $d_i$ and $u_i$ the quark singlets, $L_i$ the lepton doublets and $e_i$ the lepton singlets of the $i$-th generation. $H_u$ is the up-type Higgs doublet, $H_d$ the down-type Higgs doublet and superpartners are marked with a tilde. For instance, $\tilde{G}$ labels the gluino with mass $M_3$, $\tilde{B}$ the bino with mass $M_1$ and $\tilde{W}$ the wino with mass $M_2$.

After matching the SM with the MSSM at the SUSY scale at one-loop \cite{Carena:1999py, Buras:2002vd}, we obtain for the affected Yukawa couplings of down-type quarks and charged leptons
\begin{equation}\label{Eq:Matching}
 y_i^\mathrm{MSSM} = \frac{y_i^\mathrm{SM}}{\cos \beta \,(1 + \epsilon_i \tan \beta)},
\end{equation}
where $y_i^\mathrm{SM}=\bar{m}_i/v$ is the SM Yukawa coupling related to the $\overline{\text{MS}}$-mass $\bar{m}_i$ of the respective particle. With $v_u$ and $v_d$ being the vacuum expectation values (vevs) of the up-type and down-type Higgs doublets, $\tan \beta$ is defined as $\tan\beta=v_u/v_d$ and the vev $v$ of the SM Higgs is given by $v^2=v_u^2+v_d^2$. The quantities $\epsilon_i$ which govern the $\tan\beta$-enhanced corrections, with $i$ = $d$, $s$, $b$, $e$, $\mu$ and $\tau$, will be specified in section \ref{Sec:corrections}.
Note that since the calculation is one-loop, we can use $\overline{\text{MS}}$-quantities as well as $\overline{\text{DR}}$-quantities. 
For large $\tan \beta$ the correction $\epsilon_i \tan \beta$ can become quite large, so higher order calculations seem to be necessary, however it can be shown that the $\tan \beta$-enhanced vertex corrections are absent in higher orders \cite{Carena:1999py}. In the following, we will restrict ourselves to the case of large $\tan \beta$.  

For the up-type quarks (and neutrinos) there are no $\tan \beta$-enhanced corrections, and the remaining corrections are negligibly small. We will therefore match the up-type quark Yukawa couplings using the tree-level relation $y_i^\mathrm{MSSM} = y_i^\mathrm{SM}/\sin\beta$, with $i$ = $u$, $c$ and $t$. For the neutrino sector we will assume that the small masses are generated by the (type I) seesaw mechanism \cite{seesaw} operating at high energies close to the GUT scale.
In the following, we will drop the MSSM-label for the Yukawa couplings and imply $y_i \equiv y_i^\mathrm{MSSM}$.

\subsection{Corrections to Quark and Lepton Yukawa Couplings}\label{Sec:corrections}

We give now explicit formulae for the quantities $\epsilon_i$ in equation (\ref{Eq:Matching}), which in addition to $\tan \beta$ govern the size of the threshold corrections. Details of the calculation for the quarks can be found in \cite{Carena:1999py, Buras:2002vd}. The formulae are valid for the case of unbroken $SU(2)_L \times U(1)_Y$ symmetry, which is appropriate for the situation that the sparticle masses are in the TeV range and that we use matching conditions at these high energies above the electroweak (EW) scale $M_{\mathrm{EW}}$.

Turning to the corrections for the down-type quarks we can decompose $\epsilon_i = \epsilon_i^G + \epsilon_i^B + \epsilon_i^W + \epsilon^y \delta_{ib}$, with \cite{Freitas:2007dp}
\begin{eqnarray} \label{eq:expq}
 \epsilon_i^G & = & -\frac{2 \alpha_S}{3 \pi} \frac{\mu}{M_3} H_2(u_{\tilde{Q}_i},u_{\tilde{d}_i}) \; , \\
\label{eq:expq_bino}
 \epsilon_i^B & = & \frac{1}{16 \pi^2} \left[ \frac{{g'}^2}{6} \frac{M_1}{\mu} \left( H_2(v_{\tilde{Q}_i}, x_1) + 2 H_2(v_{\tilde{d}_i}, x_1) \right) + \frac{{g'}^2}{9} \frac{\mu}{M_1} H_2(w_{\tilde{Q}_i},w_{\tilde{d}_i}) \right], \\
 \epsilon_i^W & = & \frac{1}{16 \pi^2} \frac{3 g^2}{2} \frac{M_2}{\mu} H_2(v_{\tilde{Q}_i}, x_2) \; , \\
 \epsilon^y &=& - \frac{y_t^2}{16 \pi^2} \frac{A_t}{\mu} H_2(v_{\tilde{Q}_3}, v_{\tilde{u}_3}) \; ,
\label{eq:expq_end}
\end{eqnarray}
where $u_{\tilde{f}} = m_{\tilde{f}}^2/M_3^2$, $v_{\tilde{f}} = m_{\tilde{f}}^2/\mu^2$, $w_{\tilde{f}} = m_{\tilde{f}}^2/M_1^2$, $x_1 = M_1^2/\mu^2$ and $x_2 = M_2^2/\mu^2$ for $i=d,s,b$ and where all mass parameters are assumed to be real. The correction $\epsilon^y$ is only relevant for the $b$-quarks because of the strong hierarchy of the quark Yukawa couplings. 
The function $H_2$ is defined as
\begin{equation}
H_2(x,y) = \frac{x \ln x}{(1-x)(x-y)} + \frac{y \ln y}{(1-y)(y-x)}.
\end{equation}
Note that $H_2$ is negative for positive $x$ and $y$ and $|H_2|$ is maximal, if its arguments are minimal, and vice versa.

The corrections for the charged leptons stem from diagrams similar to the ones for the quarks and are shown in figure \ref{fig:leptons}. One difference between the corrections for quarks and charged leptons is of course that the SUSY QCD loop contributions $\epsilon_i^G$ are absent. Another difference concerns the contributions $\epsilon_i^B$ with binos in the loops, where due to the different hypercharge of the charged leptons the prefactors for these contributions are changed. In the last term in equation (\ref{eq:expl_bino}) this causes an enhancement by a factor of $-9$ compared to the corresponding term in the quark sector in equation (\ref{eq:expq_bino}). The contribution from the diagrams with winos $\epsilon_i^W$, on the other hand, is equal for quarks and leptons. A further difference between the corrections for quarks and charged leptons is that in the considered seesaw framework the $\tau$-lepton Yukawa coupling does not have a relevant correction of the $\epsilon^y$ type because the corresponding vertex correction is suppressed by the heavy mass scale of the right-handed neutrinos. For the corrections for the charged leptons we find
\begin{eqnarray} \label{eq:expl}\label{eq:expl_bino}
 \epsilon_i^B & = & \frac{1}{16 \pi^2} \left[ \frac{{g'}^2}{2} \frac{M_1}{\mu} \left(- H_2(v_{\tilde{L}_i}, x_1) + 2 H_2(v_{\tilde{e}_i}, x_1) \right) - {g'}^2 \frac{\mu}{M_1} H_2(w_{\tilde{L}_i},w_{\tilde{e}_i}) \right] , \\
 \epsilon_i^W & = & \frac{1}{16 \pi^2} \frac{3 g^2}{2} \frac{M_2}{\mu} H_2(v_{\tilde{L}_i}, x_2) \; , \label{eq:expl_end}
\end{eqnarray}
for $i=e,\mu,\tau$.

\subsection{Renormalisation Group Running from EW to the GUT Scale}

The calculation of the Yukawa couplings of the quarks and charged leptons at the GUT scale can be accomplished by solving the corresponding renormalisation group equations (RGEs) in the (type I) seesaw scenario from low-energy to the GUT scale. For this we use the package REAP introduced in \cite{Antusch:2005gp}, where also a summary of the relevant RGEs can be found.

For our analysis, we take as input values the running quark and lepton masses at the top scale $\bar{m}_t (\bar{m}_t)$, which have been calculated recently in \cite{Xing:2007fb} with up-to-date experimental values for the low-energy quark and charged lepton masses, and evolve them first to the SUSY scale $M_{\mathrm{SUSY}}$ using the SM RGEs. At the SUSY scale we match the SM with the MSSM to obtain the running $\overline{\text{MS}}$ Yukawa couplings via equation (\ref{Eq:Matching}). Since we consider one-loop running, we can neglect issues of scheme dependence such as transformations from $\overline{\text{MS}}$ to $\overline{\text{DR}}$ quantities. Two-loop running (and scheme-dependent) effects are small compared to the $\tan \beta$-enhanced threshold corrections and can be neglected. 

As the next step, we solve the RGEs from the SUSY scale to $M_{\mathrm{GUT}}$ taking into account possible intermediate right-handed neutrino thresholds as discussed in \cite{Antusch:2002rr}. For our numerical calculations we use REAP, which solves the complete set of one-loop RGEs and automatically includes the right-handed neutrino thresholds.
We will comment on the possible effects of right-handed neutrino thresholds, which depend on the additional degrees of freedom in seesaw models, in section \ref{Sec:RHnus}. If not stated otherwise, they are ignored in our analysis.

We note that there are SUSY scenarios which may lead to corrections to our approach of one-step matching at the SUSY scale in the EW unbroken phase. For example, if the sparticle spectrum is light, effects of EW symmetry breaking may have to be taken into account for the calculation of the SUSY threshold corrections. Another example is the possibility of having a (relatively) split sparticle spectrum, in which case matching at one scale would be a bad approximation. When we present explicit examples in the following, we choose parameters where our assumptions are justified to a good approximation.

\section{Quark and Lepton Yukawa Couplings at the GUT Scale}

In this part we present our analytical and numerical results. We start with a semianalytic discussion of the individual contributions to the threshold corrections presented in equations (\ref{eq:expq}) to (\ref{eq:expl_end}), which allows to estimate their size as well as the size of the total corrections $\epsilon_i$. We will then turn to the numerical analysis, where we quantitatively discuss the effect of the most relevant parameters and present our final results for the quark and charged lepton Yukawa couplings (and mass ratios\footnote{We note that when we refer to fermion masses at the GUT scale, what we mean is simply the Yukawa coupling multiplied by the low-energy value of the corresponding Higgs vev. }) at the GUT scale. To isolate the effects of the SUSY (MSSM) parameters from the uncertainties induced by the quark mass errors, we first use best-fit values for the low-energy fermion masses. These errors are later included in our final results presented in tables \ref{tab:Ratfinal} and \ref{tab:Yukfinal}. We note that the quark mass errors have a significant effect, whereas the charged lepton mass errors are negligibly small.

\begin{table}
\begin{center}
\begin{tabular}{|c|c|c|c|}\hline
SUSY parameter & Case $g_+$ & Case $g_-$ & Case $a$ \\ \hline
\hline $m_{\tilde{f}}$ in TeV		& [0.5, 1.5]  & [0.5, 1.5] & [0.5, 1.5]  \\ 
\hline $M_1$ in TeV 			& [0.5, 1]    & [0.5, 1]   & [1.65, 3.3] \\ 
\hline $M_2$ in TeV 			& [1, 2]      & [1, 2]     & [0.5, 1]  \\ 
\hline $M_3$ in TeV 			& [3, 6]      & [3, 6]     & [-9, -4.5]  \\
\hline $\mu$ in TeV 			& 0.5      & -0.5    &  0.5  \\
\hline $A_t$ in TeV 			& $\pm$1  & $\pm$1 &  $\pm$1  \\
\hline $M_{\mathrm{SUSY}}$ in TeV 	& 1  & 1 &  1  \\ \hline
\end{tabular}
\end{center}
\caption{Example ranges of SUSY parameters (at the matching scale $M_{\mathrm{SUSY}}$) used in our analyses in sections \ref{Sec:Analytic} and \ref{Sec:Scan}. The choices of gaugino masses in the cases $g_\pm$ are inspired by universal gaugino masses at the GUT scale and in case $a$  by anomaly-mediated SUSY breaking. We therefore use the low-energy approximation $M_1:M_2:M_3 = 1:2:6$ for the cases $g_\pm$ and $M_1:M_2:M_3 = 3.3:1:-9$ for case $a$ as a constraint.\label{tab:SUSYpar}}
\end{table}

\subsection{Semianalytical Approach}\label{Sec:Analytic}

For our semianalytical approach we first define three example choices of possible ranges of the relevant SUSY breaking parameters. They are listed in table \ref{tab:SUSYpar} and will be referred to as cases $g_+$, $g_-$ and $a$. The cases $g_+$ and $g_-$ are inspired by scenarios with universal boundary conditions for the gauginos (where $M_1/g_1^2 = M_2/g_2^2 = M_3/g_3^2$) and case $a$ by anomaly-mediated SUSY breaking \cite{Randall:1998uk, Giudice:1998xp} (where $M_1/(g_1^2 \beta_1) = M_2/(g_2^2 \beta_2) = M_3/(g_3^2 \beta_3)$, with $(\beta_1,\beta_2, \beta_3) = (33/5 , 1, -3)$). We note that instead of these relations at the SUSY scale, we use the low-energy approximation $M_1:M_2:M_3 \approx 1:2:6$ for the cases $g_\pm$ and $M_1:M_2:M_3 \approx 3.3:1:-9$ for case $a$.
We furthermore only introduce relations for the gaugino masses, whereas for the sfermion masses $m_{\tilde{f}}$ and the $\mu$ and $A_t$ parameters we do not apply restrictions from a specific model of SUSY breaking.
The example ranges are chosen such that we can neglect effects which are suppressed by $M_{\mathrm{EW}}/M_{\mathrm{SUSY}}$, such as mixing between left- and right-handed sfermions, and that our approach of one-step matching is justified to a good approximation. 
We note that left-right mixing effects may nevertheless be important \cite{Buras:2002vd}, for instance in so-called inverted scalar mass hierarchy (ISMH) scenarios. 
We also note that without specifying the remaining SUSY parameters (which do not enter the formulae for the threshold corrections) we cannot apply various relevant phenomenological constraints on the spectrum. To do so, we would have to impose further constraints on the soft breaking parameters (as done e.g.\ in \cite{Altmannshofer:2008vr}) which is, however, not our intention.

For the parameter ranges specified in table \ref{tab:SUSYpar}, we can estimate the corresponding ranges for the different components of $\epsilon_i$. Scanning over the ranges of SUSY parameters in table \ref{tab:SUSYpar} we obtain the resulting ranges presented in table \ref{tab:analytical}. The gauge couplings are evaluated at the SUSY scale. From these ranges we can already see that the electroweak contributions cannot be neglected. In our scan we find that they can amount up to about 50~\% of the QCD contribution, larger than one might suspect from table \ref{tab:SUSYpar} due to correlations between the corrections. 
We can also see from table \ref{tab:analytical} that, for the quarks, in case $a$ the inclusion of the electroweak corrections results in an enlargement of the threshold correction because here the QCD and the electroweak corrections add up, whereas in case $g_\pm$ it results in a reduction of the total correction because both contributions partially cancel.

\begin{table}
\begin{center}
\begin{tabular}{|c|c|c|c|}\hline
						&   Case $g_+$   & Case $g_-$ & Case $a$ \\ \hline
\hline $\epsilon_i^G$ in $10^{-3}$ 		&   [3.52, 9.31] & [-9.31, -3.52] & [-7.85, -3.10] \\ 
\hline $\epsilon_i^B$ for quarks in $10^{-3}$ 	& [-0.31, -0.08] &   [0.08, 0.31] & [-0.30, -0.16] \\
\hline $\epsilon_i^B$ for leptons in $10^{-3}$ 	& [-0.18,  0.30] & [-0.30,  0.18] &  [0.01,  0.30]  \\  
\hline $\epsilon_i^W$ in $10^{-3}$ 		& [-2.21, -0.98] &   [0.98, 2.21] & [-2.21, -0.72] \\ 
\hline $\epsilon^y$  in $10^{-3}$  		& sign($A_t$) [0.84, 4.65] & -sign($A_t$) [0.84, 4.65] & sign($A_t$) [0.84, 4.65] \\ \hline
\end{tabular}
\end{center}
\caption{Ranges for the various contributions to the SUSY threshold corrections (corresponding to the example ranges of SUSY parameters in table \ref{tab:SUSYpar}) for $i=d,s,b$ and $i=e,\mu,\tau$, respectively. For the charged leptons there is no contribution $\epsilon_i^G$ and $\epsilon^y$. The ranges for $\epsilon_i^W$ are the same for quarks and leptons.\label{tab:analytical} }
\end{table}

The ranges for the $\epsilon_i$ from table \ref{tab:analytical} can now be used to obtain (naive) analytic estimates for the ratios of the fermion masses (or Yukawa couplings) at the GUT scale. For example, in leading order the GUT scale ratio $m_e/m_d$ is given by
\begin{equation}
 \frac{m_e(M_{\mathrm{GUT}})}{m_d(M_{\mathrm{GUT}})} \approx \frac{\hat{m}_e(M_{\mathrm{GUT}})}{\hat{m}_d(M_{\mathrm{GUT}})} \:\frac{1 + \epsilon_d \tan \beta}{1 + \epsilon_e \tan \beta} = \frac{\hat{m}_e(M_{\mathrm{GUT}})}{\hat{m}_d(M_{\mathrm{GUT}})} \left( 1 + \left(\epsilon_d - \epsilon_e \right) \tan \beta \right) + \mathcal{O}(\epsilon_e^2 \tan^2 \beta),
\end{equation}
where $\hat{m}(M_{\mathrm{GUT}})$ denotes the fermion masses at the GUT scale without SUSY thresholds. We will use the analogous formula for the second generation. For the third generation we take the ratios of the Yukawa couplings so we have to take into account an additional $\tan \beta$ factor for $y_t/y_b$. 
We will later on compare these estimates with the numerical results for the same ranges of MSSM parameters (cf.~table (\ref{tab:serror})). For the values of the masses and Yukawa couplings at the GUT scale, we take the values calculated with REAP setting all SUSY threshold corrections to zero and using the best-fit values for the fermion masses as low-energy input. These values for the Yukawa couplings are collected in table \ref{tab:Yuknothresholds}. In the following (e.g. in table \ref{tab:GJRana}) we will refer to the case without SUSY thresholds as case 0.

\begin{table}
\begin{center}
\begin{tabular}{|c|c|c|c|}\hline
 &  $\hat{y}_e$ in $10^{-4}$ & $\hat{y}_\mu$ in $10^{-2}$ & $\hat{y}_\tau$  \\ \hline \hline
$\tan \beta = 30$ & 0.62 & 1.31 & 0.23 \\ \hline
$\tan \beta = 40$ & 0.88 & 1.85 & 0.34 \\ \hline
$\tan \beta = 50$ & 1.21 & 2.55 & 0.51 \\ \hline
\multicolumn{1}{c}{}
\end{tabular}

\begin{tabular}{|c|c|c|c|}\hline
 &  $\hat{y}_d$ in $10^{-4}$ & $\hat{y}_s$ in $10^{-2}$ & $\hat{y}_b$  \\ \hline \hline
$\tan \beta = 30$ & 1.57 & 0.30 & 0.18 \\ \hline
$\tan \beta = 40$ & 2.22 & 0.43 & 0.26 \\ \hline
$\tan \beta = 50$ & 3.06 & 0.59 & 0.39 \\ \hline
\multicolumn{1}{c}{}
\end{tabular}

\begin{tabular}{|c|c|c|c|}\hline
 &  $\hat{y}_u$ in $10^{-6}$ & $\hat{y}_c$ in $10^{-4}$ & $\hat{y}_t$ \\ \hline \hline
$\tan \beta = 30$ & 2.73 & 1.33 & 0.49 \\ \hline
$\tan \beta = 40$ & 2.75 & 1.34 & 0.50 \\ \hline
$\tan \beta = 50$ & 2.77 & 1.35 & 0.52 \\ \hline
\end{tabular}

\end{center}
\caption{Best-fit values for the Yukawa couplings at the GUT scale without SUSY threshold corrections (case 0) for $M_{\mathrm{SUSY}} = 1$~TeV and different values of $\tan \beta$.\label{tab:Yuknothresholds} }
\end{table}

\begin{table}
\begin{center}
\begin{tabular}{|c|c|c|c|c|}\hline
			& Case 0 & Case $g_+$ & Case $g_-$  & Case $a$ \\ \hline
\hline $m_e/m_d$ 	& 0.39 & [0.35, 0.64] & [0.15, 0.44] & [0.16, 0.44] \\ 
\hline $m_\mu/m_s$ 	& 4.35 & [3.83, 7.01] & [1.69, 4.87] & [1.81, 4.85] \\ 
\hline $y_\tau/y_b$ 	& 1.32 & [1.16, 2.13] & $\leq 1.48$  & $\leq 1.47$ \\ 
\hline $y_t/y_b$ 	& 1.93 & [1.65, 2.92] & $\leq 2.21$  & $\leq 1.98$ \\ \hline
\end{tabular}
\end{center}
\caption{Semi-analytic (naive) estimates for the ranges of the mass and Yukawa coupling ratios at the GUT scale (corresponding to the example ranges of SUSY parameters in table \ref{tab:SUSYpar}) for $\tan \beta = 40$. Case 0 refers to the case without SUSY threshold corrections. For the ranges involving $y_b$, the lower boundaries depend on the cut we had to introduce in order to keep $y_b$ perturbative up to $M_{\mathrm{GUT}}$ and therefore have been omitted. \label{tab:GJRana}}
\end{table}

The results of these estimates are collected in table \ref{tab:GJRana}. We note that these estimates are naive in the sense that we have combined the maximal and minimal values of each of the contributions to $\epsilon_i$, neglecting possible correlations between them. For example, we do not account for the effect that the QCD corrections become large, if $M_3$ and thus the gaugino masses are large and the sfermion masses small, whereas the wino corrections become large, if the gaugino masses and the sfermion masses are small. However, as we will later see, the estimates nevertheless work surprisingly well. 
Another effect which we can immediately see from the analytic estimates is that $y_b$ can become nonperturbatively large if $\epsilon_b \tan \beta$ is large and negative. In fact, it can even occur that $\epsilon_i \tan \beta \leq -1$, which spoils the perturbative expansion.
Wherever non-perturbative values of $y_b$ occur, we only give the upper boundaries of the ranges $y_\tau/y_b$ and $y_t/y_b$ and the lower boundary for $y_b$ itself.

The naive estimates already suggest that with SUSY thresholds included, a wide range of GUT scale values of down-type quark and charged lepton masses (or Yukawa couplings) could be realised. Taking a preliminary look at the predicted ratio for $m_\mu/m_s$, the naive estimates suggest that with the SUSY parameters of example $g_+$, the GUT scale value of $m_\mu/m_s$ is typically significantly larger than the GJ relation of $m_\mu/m_s = 3$. On the other hand, scenario $g_-$ and $a$ are well compatible with the GJ relation. Beyond the GJ relation, the naive estimates also imply that with SUSY thresholds included, a large variety of GUT model predictions for these ratios might be compatible with the low-energy data on quark and lepton masses. A full numerical analysis for the example SUSY parameter ranges $g_\pm$ and $a$ will be presented in section \ref{Sec:Scan}.

\subsection{Dependence on $\boldsymbol{\mu}$ and $\boldsymbol{\tan \beta}$}

\begin{figure}
 \centering
 \includegraphics[scale=0.6]{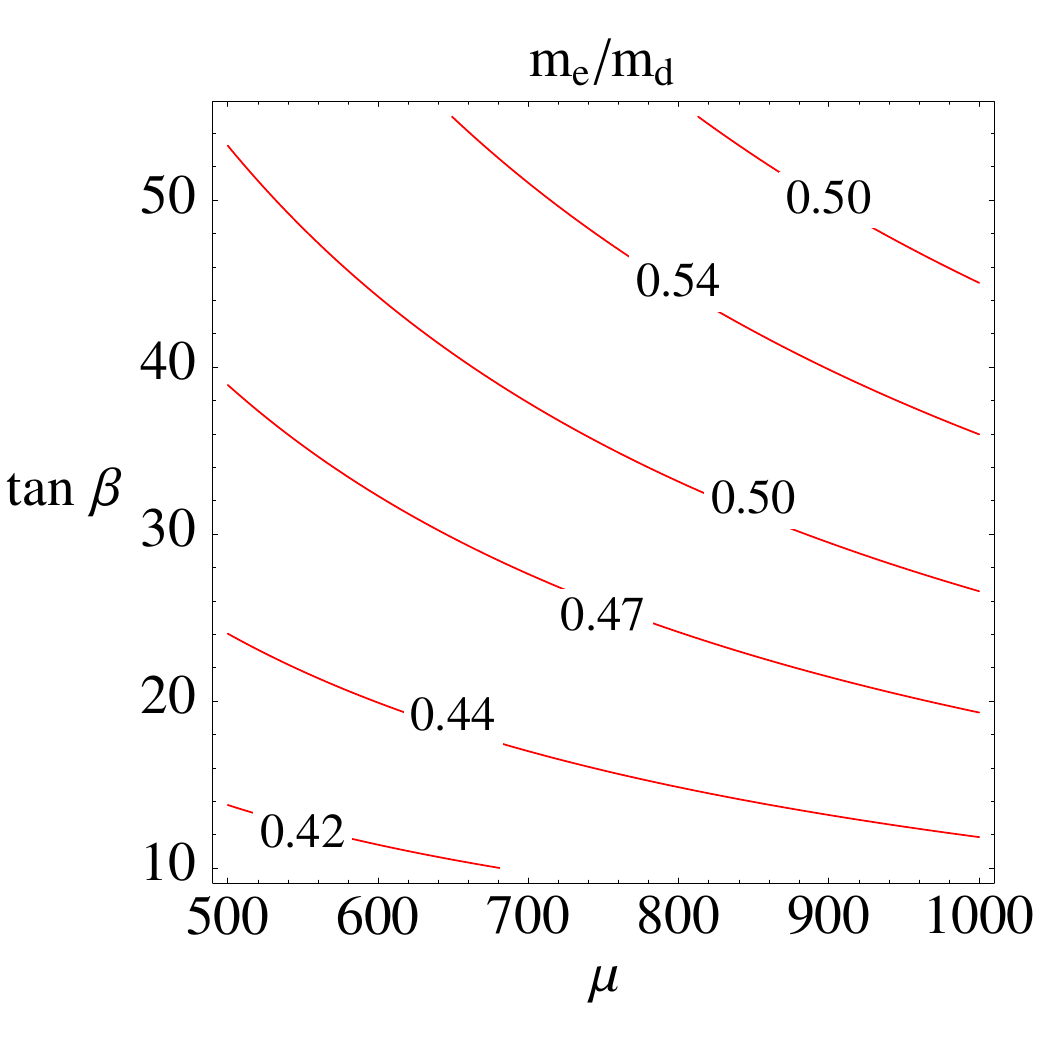}
 \includegraphics[scale=0.6]{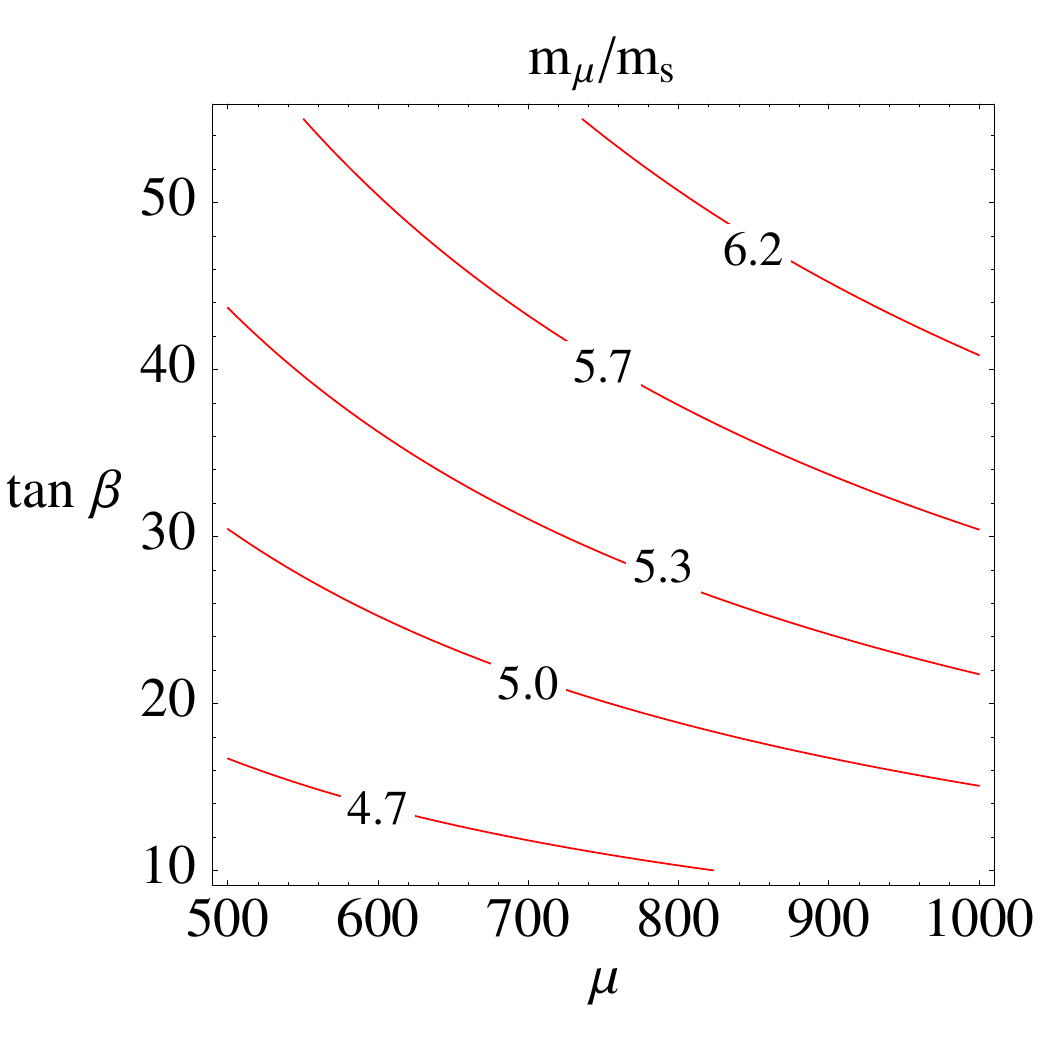}
 \includegraphics[scale=0.6]{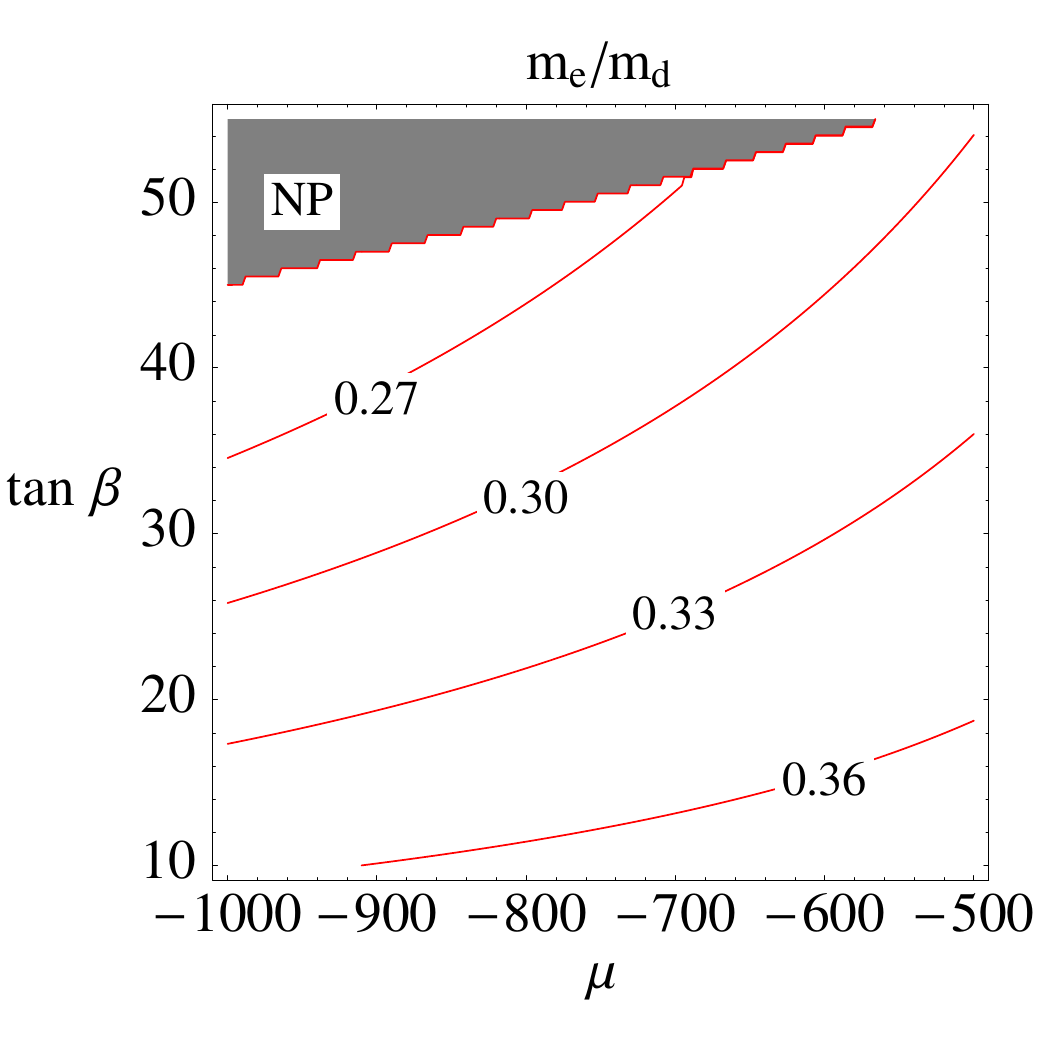}
 \includegraphics[scale=0.6]{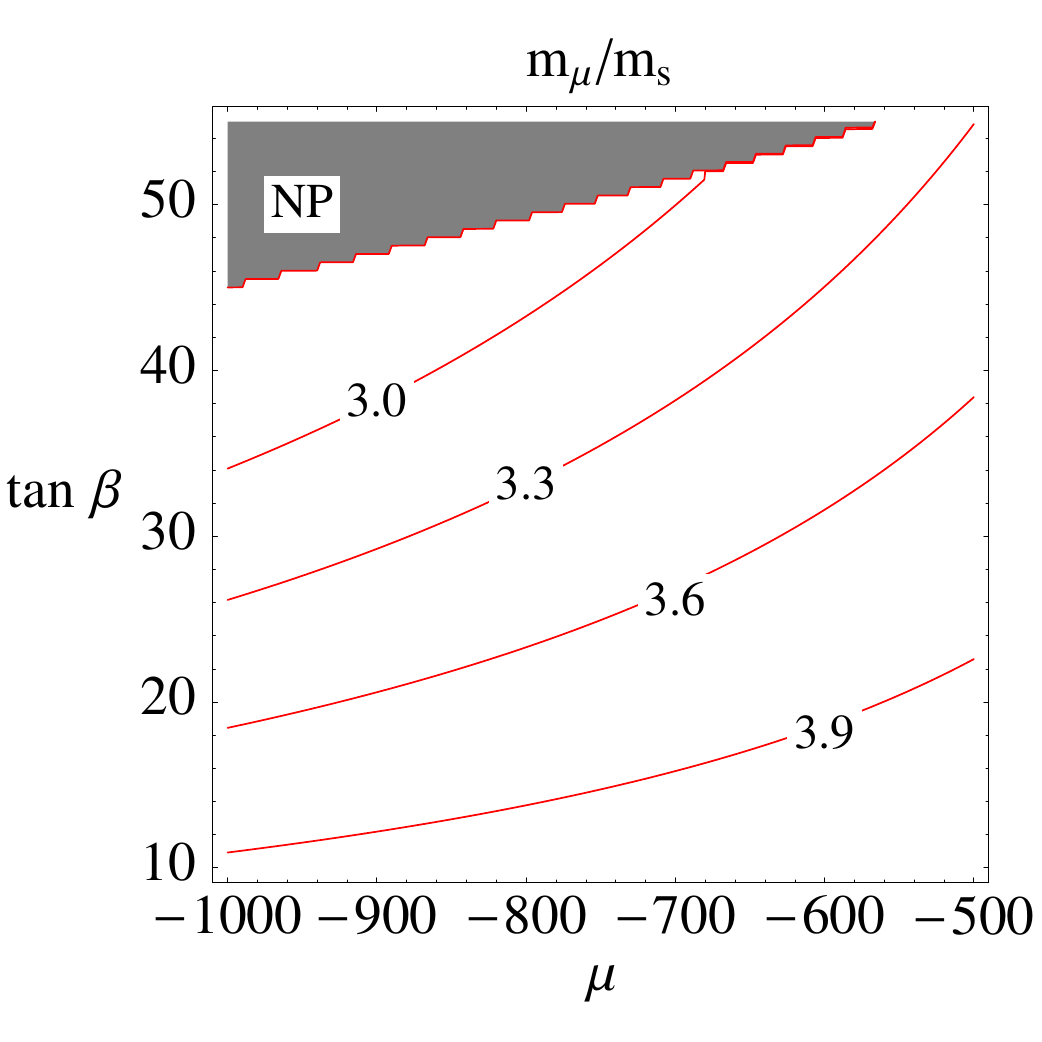}
 \includegraphics[scale=0.6]{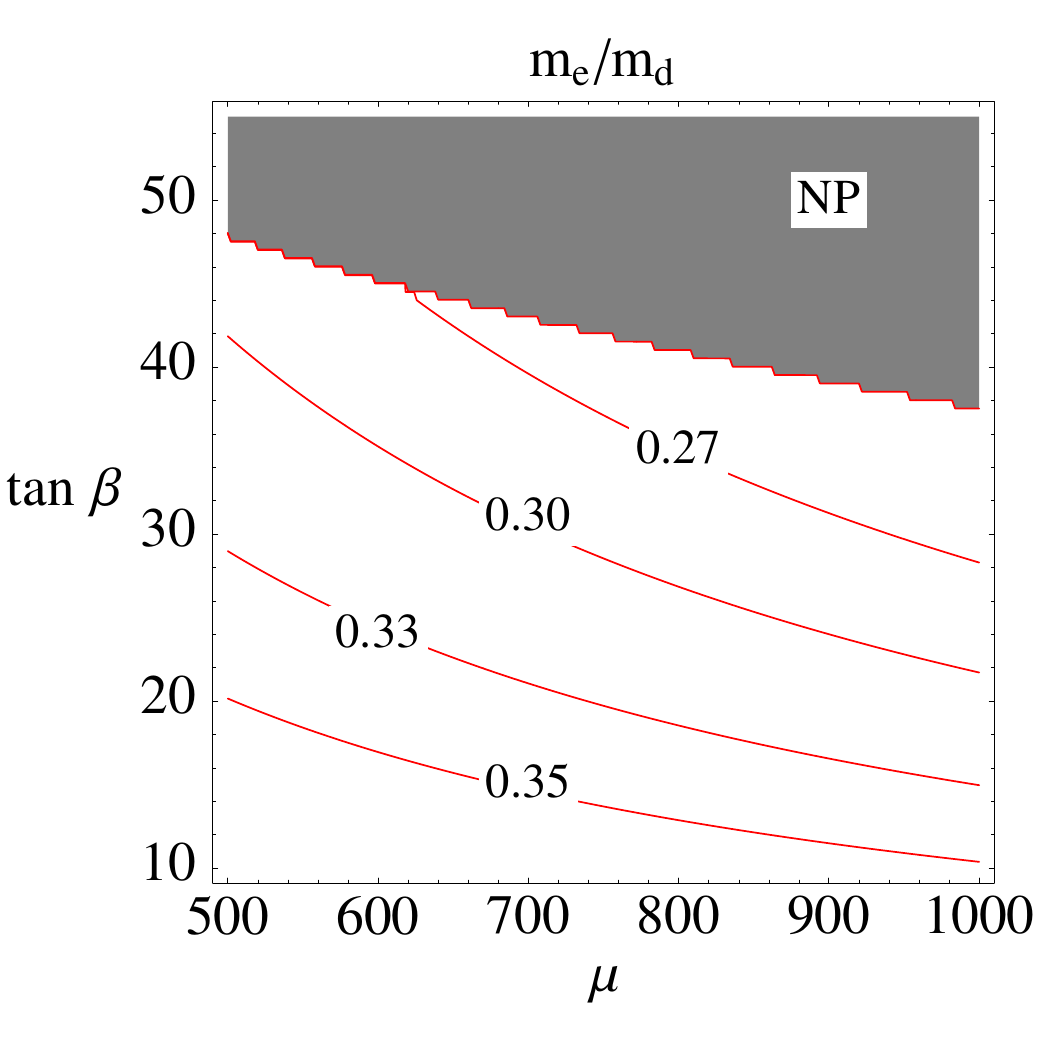}
 \includegraphics[scale=0.6]{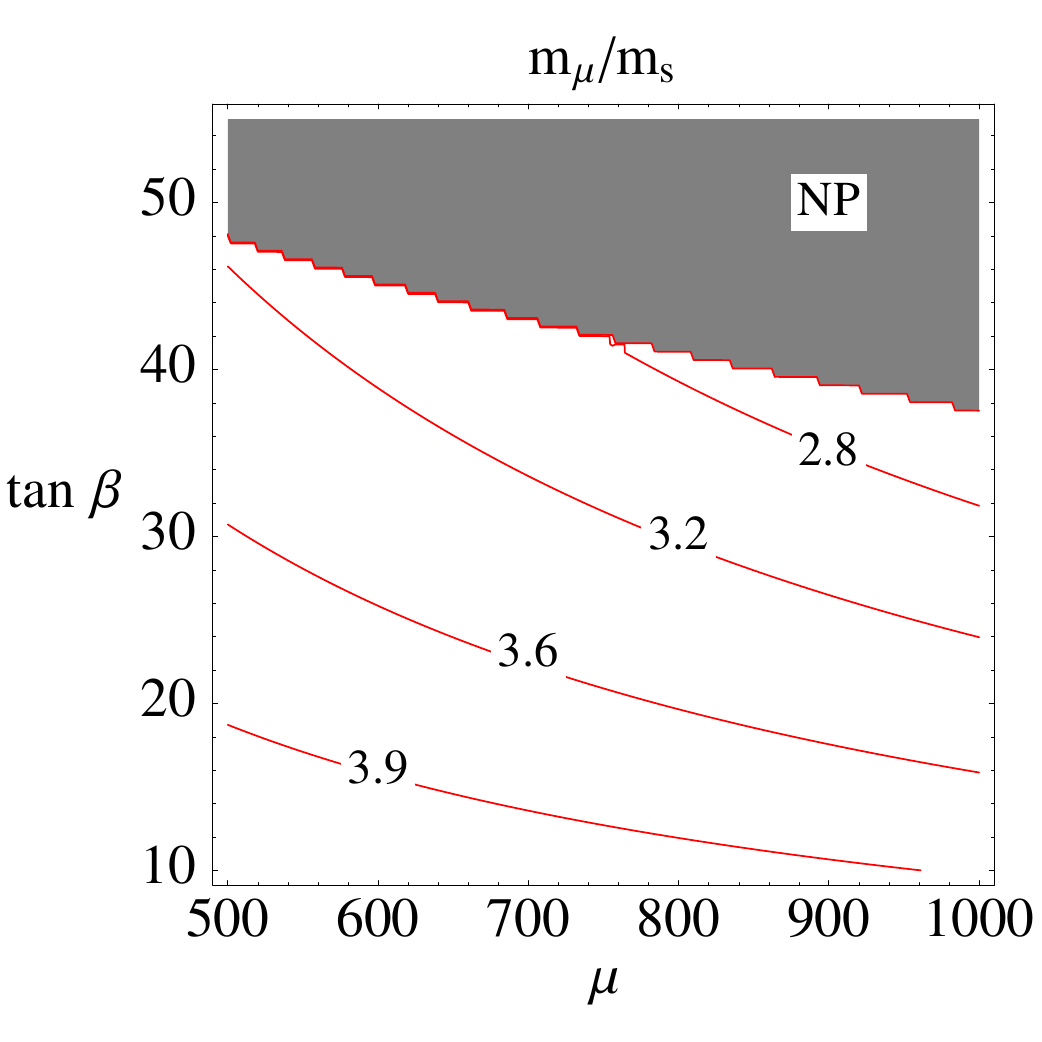}
 \caption{Contour plots for the GUT scale ratios $m_e/m_d$ (left side) and $m_\mu/m_s$ (right side) for $M_3>0$, $\mu>0$ (first row), $M_3>0$, $\mu<0$ (second row) and $M_3<0$, $\mu>0$ (third row) in the $\mu$-$\tan \beta$ plane. In the grey areas labeled with NP the value of $y_b$ becomes nonperturbatively large. \label{fig:Contour12}}
\end{figure}

\begin{figure}
 \centering
 \includegraphics[scale=0.6]{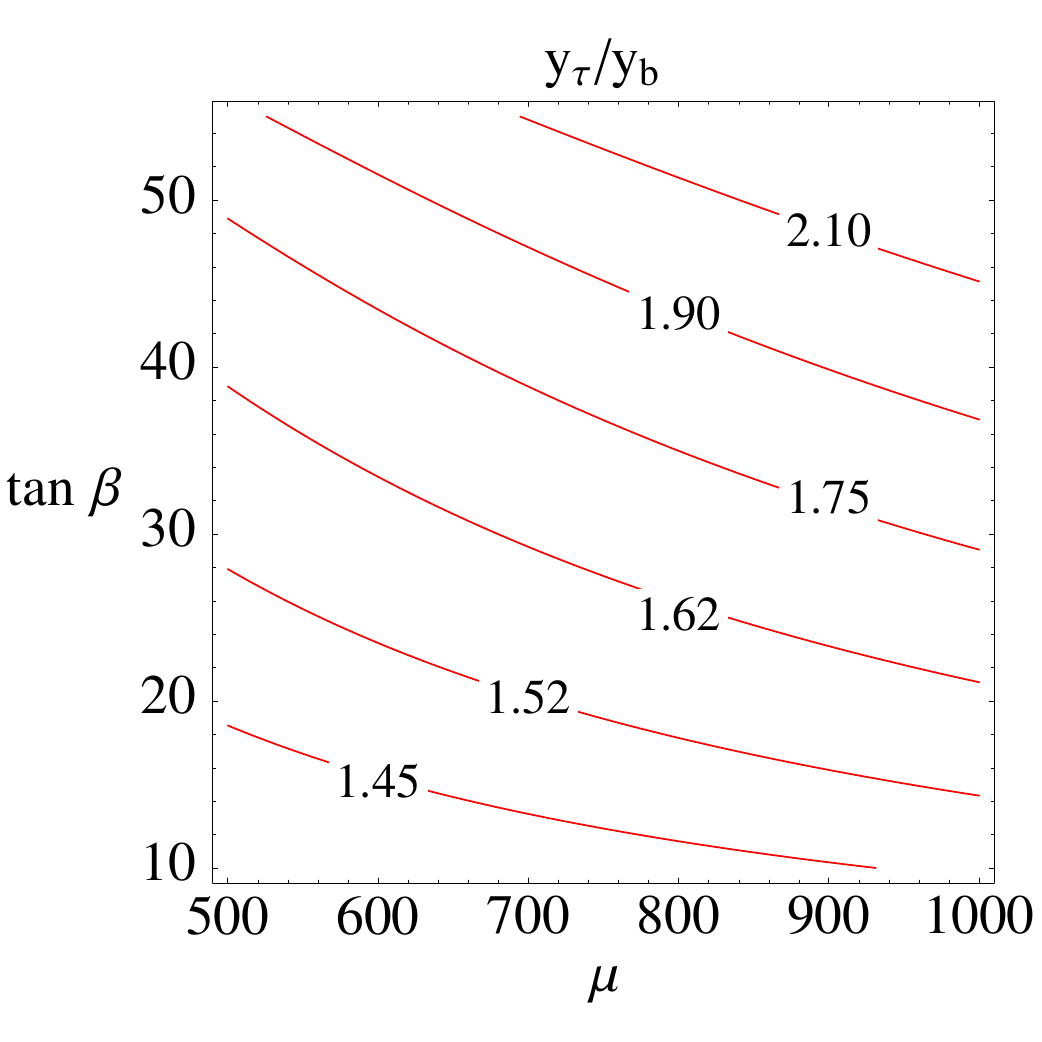}
 \includegraphics[scale=0.6]{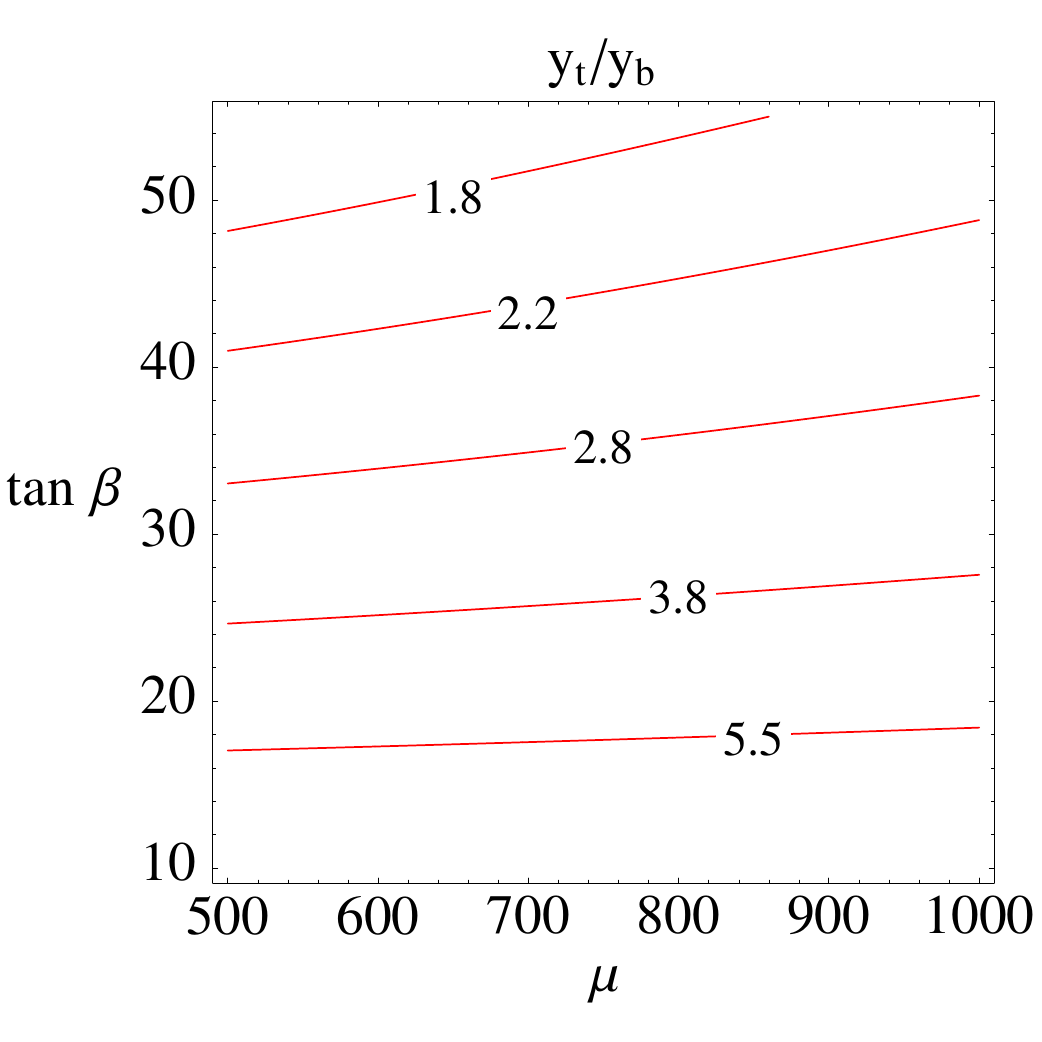}
 \includegraphics[scale=0.6]{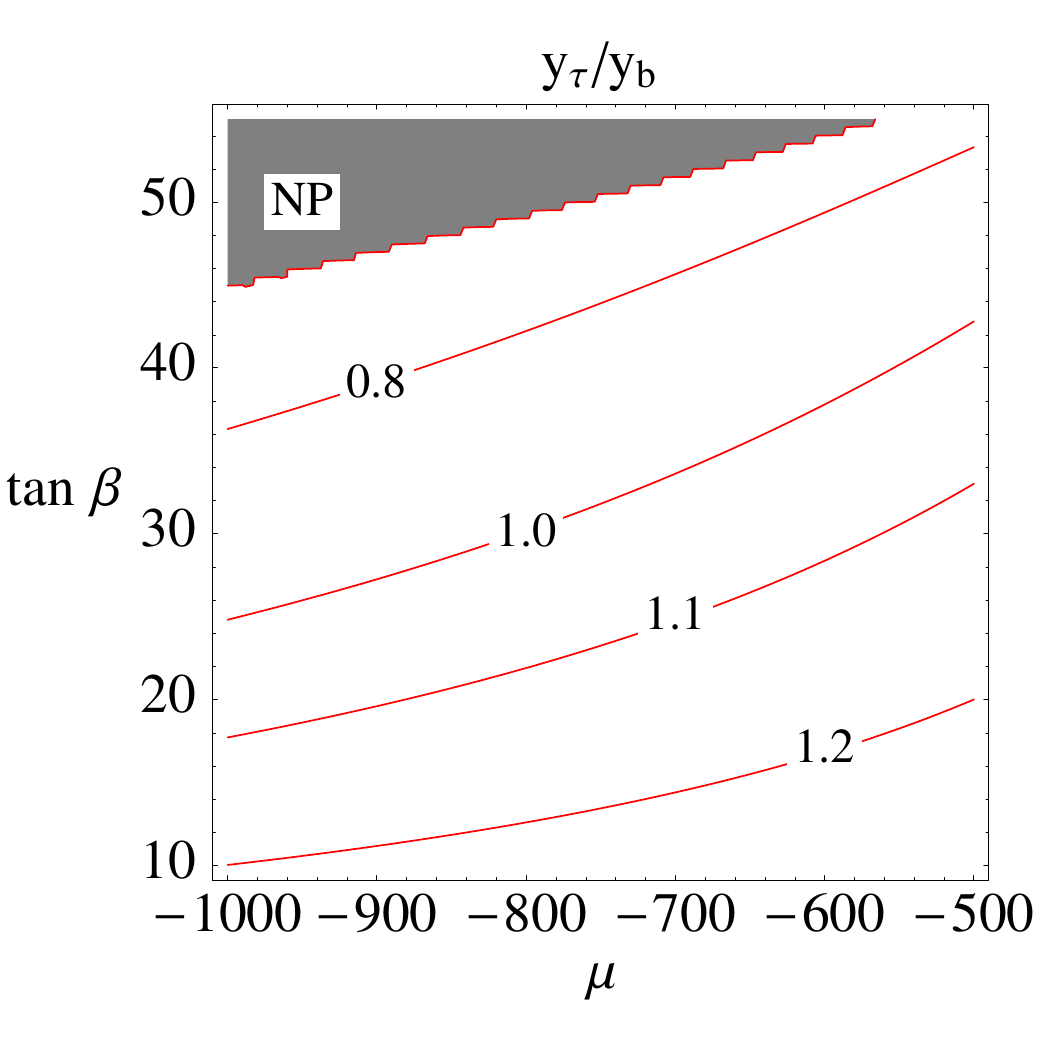}
 \includegraphics[scale=0.6]{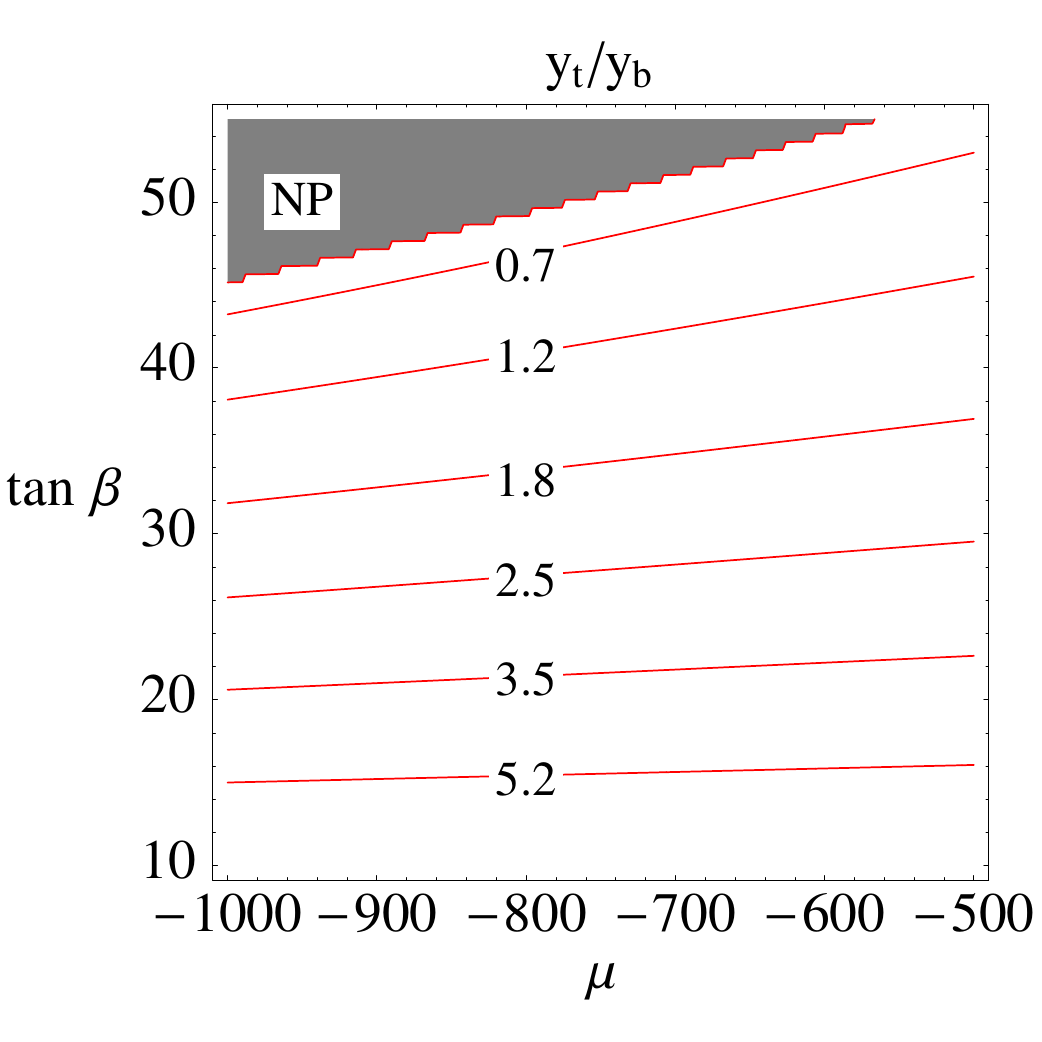}
 \includegraphics[scale=0.6]{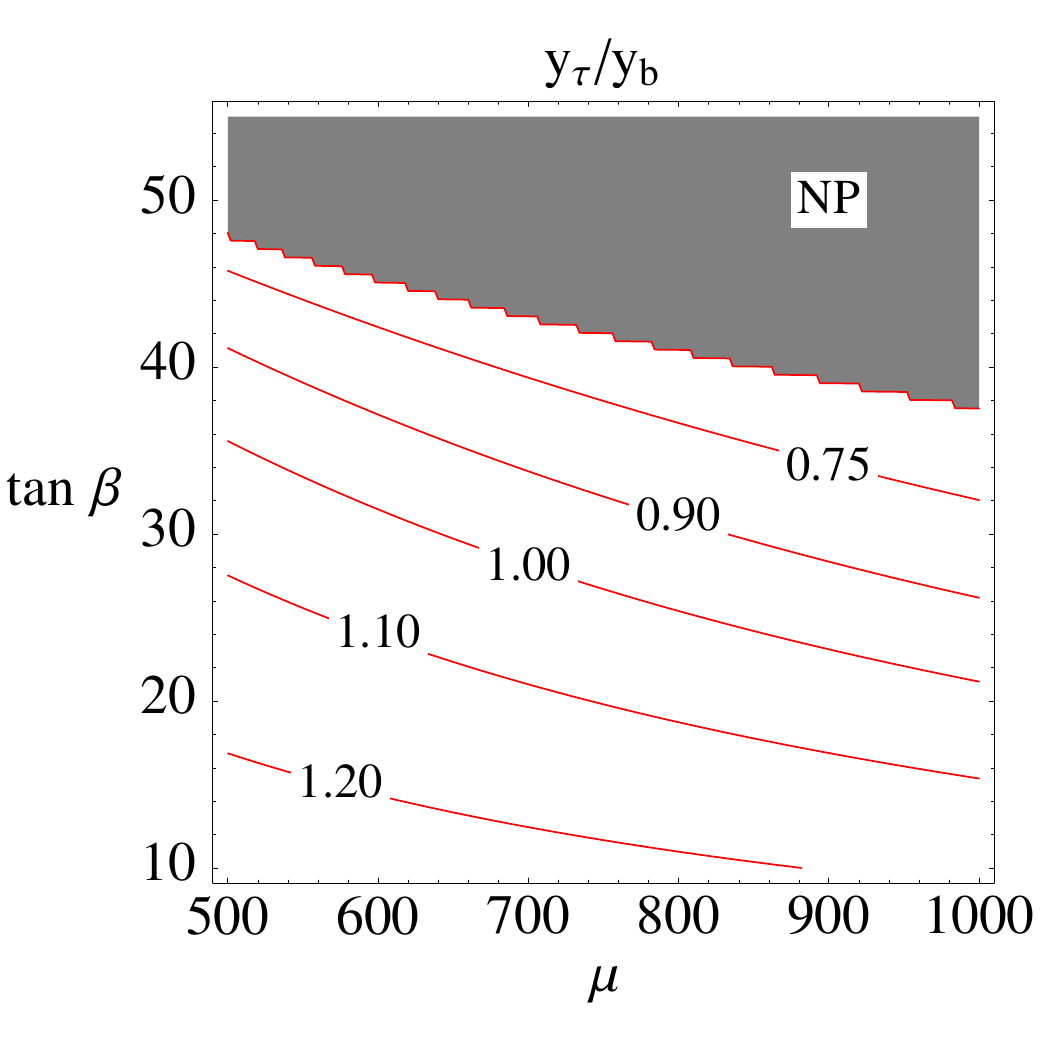}
 \includegraphics[scale=0.6]{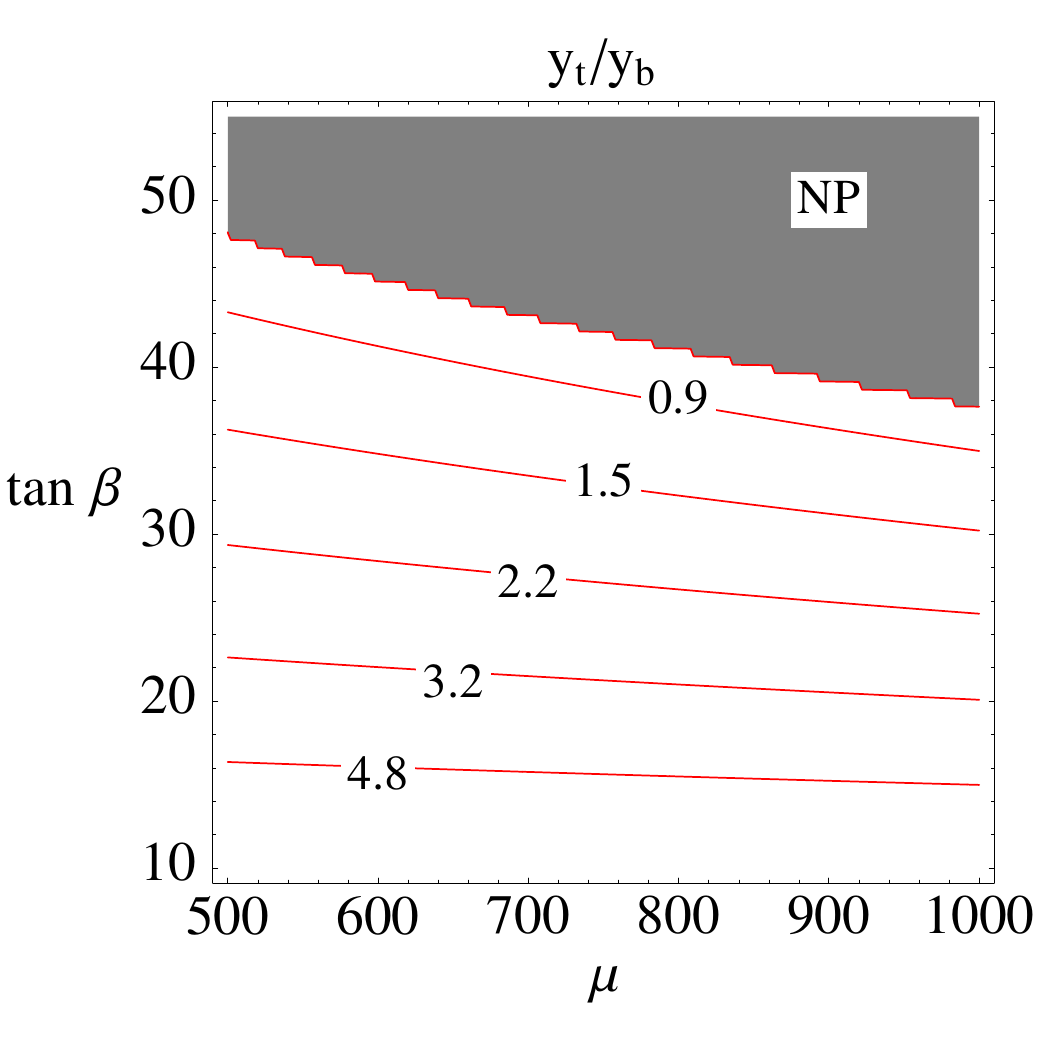}
 \caption{Contour plots for the GUT scale ratios $y_\tau/y_b$ (left side) and $y_t/y_b$ (right side) for $M_3>0$, $\mu>0$ (first row), $M_3>0$, $\mu<0$ (second row) and $M_3<0$, $\mu>0$ (third row) in the $\mu$-$\tan \beta$ plane. In the grey areas labeled with NP the value of $y_b$ becomes nonperturbatively large. \label{fig:Contour34}}
\end{figure}

Before we proceed with the numerical analysis for the example ranges of MSSM para\-meters of table \ref{tab:SUSYpar}, we discuss the dependence on $\mu$ and $\tan \beta$, which has been kept fixed in the last section. The dependence on $\mu$ is rather important, because all corrections are proportional to $\mu$ or $1/\mu$. The parameter $\mu$ therefore gives the overall sign of the corrections and determines if the Yukawa couplings are enhanced or reduced by the SUSY threshold effects. In addition, $\tan \beta$ is very important because the threshold corrections are almost linear in $\tan \beta$ and also because for successful third family Yukawa unification we need a large value of $\tan \beta$.
To isolate the effects of these parameters, we have effectively turned off the right-handed neutrino 
threshold effects, put $A_t$ to zero and all the other soft SUSY breaking parameters and the SUSY scale to $1$ TeV (with both signs allowed for $M_3$ but with $M_1>0,M_2>0$). 
In figures \ref{fig:Contour12} and \ref{fig:Contour34} the numerical results are presented as contour plots in the $\mu$-$\tan \beta$ plane for the four ratios $m_e/m_d$, $m_\mu/m_s$, $y_\tau/y_b$ and $y_t/y_b$ for different combinations of the sign of $\mu$ and $M_3$. 

There are several interesting points we would like to remark on: First, the overall dependence in the plots illustrates the anticipated behaviour from the fact that the leading contribution from gluino loops is proportional to $\mu$ and that the overall size of the corrections is proportional to $\tan \beta$. They also illustrate that for $\mu M_3 < 0$ (second and third rows in the figures) the total corrections enhance the down-type quark Yukawa couplings leading to more stringent restrictions for the possible values of $\tan \beta$ from perturbativity of $y_b$ up to the GUT scale. On the other hand, for $\mu M_3 > 0$ and $M_3 > 0$ (first row in the figures) the total corrections lower the down-type quark masses and in principle larger values for $\tan \beta$ are possible. Second, interesting conclusions can also be drawn from comparing the second to the third row of the figures. From the leading SUSY QCD contribution which is invariant under a simultaneous change of sign in $\mu$ and $M_3$, one might expect that the plots in the second and third rows look very similar (if understood as results for $|\mu|$). Differences are entirely  induced by the contributions from wino and bino loops, since we have chosen $A_t = 0$. Inspecting the numerical results shows significant differences in the results for $\mu < 0$ and $M_3 > 0$ and $\mu > 0$ and $M_3 < 0$, which confirms that the EW contributions are indeed important and cannot be ignored (as we have already concluded from our semianalytic estimates in section \ref{Sec:Analytic}).

\subsection{Dependence on $\boldsymbol{M_\mathrm{SUSY}}$ and $\boldsymbol{A_t}$}

The GUT scale values of the quark and lepton Yukawa couplings also depend on $M_\mathrm{SUSY}$ and $A_t$. While the correction to the bottom Yukawa coupling can be significant (as can be seen from table \ref{tab:analytical}), the effects on the down and strange quark Yukawa couplings are quite weak since they  only stem from indirect effects (modified RG evolution) due to the change of the bottom Yukawa coupling.

We also looked explicitly at the dependence on $M_\mathrm{SUSY}$ by fixing all other parameters and varying only $M_\mathrm{SUSY}$. We found that changing $M_\mathrm{SUSY}$ can have some effect on the GUT scale value of the Yukawa couplings due to the difference in the RGEs between SM and MSSM; however this effect is typically much smaller than the uncertainty induced by the quark mass errors and the sparticle spectrum, and it nearly cancels out when we consider ratios of masses or Yukawa couplings. We have therefore fixed $M_\mathrm{SUSY}$ to $1$ TeV in our numerical examples.

\subsection{Right-handed Neutrino Threshold Effects}\label{Sec:RHnus}

For analysing the possible dependence on threshold effects from the right-handed neutrino sector, we have taken the three examples for SUSY parameters from our analysis on the $\mu$ and $\tan \beta$ dependence, but fixed $\mu$ to $\pm 0.5$ TeV and $\tan \beta$ to 40. For these example parameter points we have investigated the effects for the three different scenarios of sequential dominance \cite{SD} also used as examples in \cite{Antusch:2007dj}. With largest neutrino Yukawa couplings being ${\cal O}(1)$, we found that the deviations are typically smaller than 5~\%, which is small compared to the SUSY threshold effects (in our examples) and also compared to the uncertainties induced by the present quark mass errors, especially for the first and second generation.  

\subsection{Impact of the Sparticle Spectrum}\label{Sec:Scan}

As already stated in the previous sections, the GUT scale values of the quark and lepton Yukawa couplings (and masses) strongly depend on the sparticle spectrum. We will now analyse its impact numerically in more detail. Because of the large number of relevant parameters, we do not attempt to discuss each of them separately, but rather make a parameter scan.

For our scan, we take the three example ranges of SUSY parameters used for our analytical estimates in section \ref{Sec:Analytic}, which are listed explicitly in table \ref{tab:SUSYpar}. In addition, for $\tan \beta$ we assume a range from 30 to 50. The sfermion mass parameters and $A_t$ are scanned with a step size of 1 TeV (including $A_t = 0$), the mass of the lightest gaugino was changed with a step size of 0.5 TeV and $\tan \beta$ is scanned  with a step size of 10. 
Although this seems to be a rather coarse scan, we note that we have shown in section \ref{Sec:Analytic} that the extremal values of the SUSY threshold corrections correspond to the extremal values of the SUSY parameters. Therefore, increasing the number of plotted points would not lead to enlarged ranges for the GUT scale quantities. 
The parameter points for which $y_b$ becomes non-perturbative have been dropped from our analysis. We note that for simplicity we have introduced a slightly more restrictive cut and included only parameter points where $y_b < 1$ at $M_\mathrm{SUSY}$ (which ensures perturbativity up to $M_\mathrm{GUT}$ but eventually removes a few allowed parameter points with large but still perturbative $y_b$). 
Furthermore we have chosen $M_{\mathrm{SUSY}}$ to be 1~TeV. The masses of the first two sfermion generations have been assumed identical, which is inspired by universal high scale boundary conditions for sfermions. We note that a small mass splitting does not change the conclusions from this plot; however a large mass splitting can reduce the threshold effects due to a reduced value of the function $|H_2|$ in the formulae (\ref{eq:expq}) to (\ref{eq:expl_end}).  

\begin{figure}
   \centering
   \includegraphics[scale=0.58]{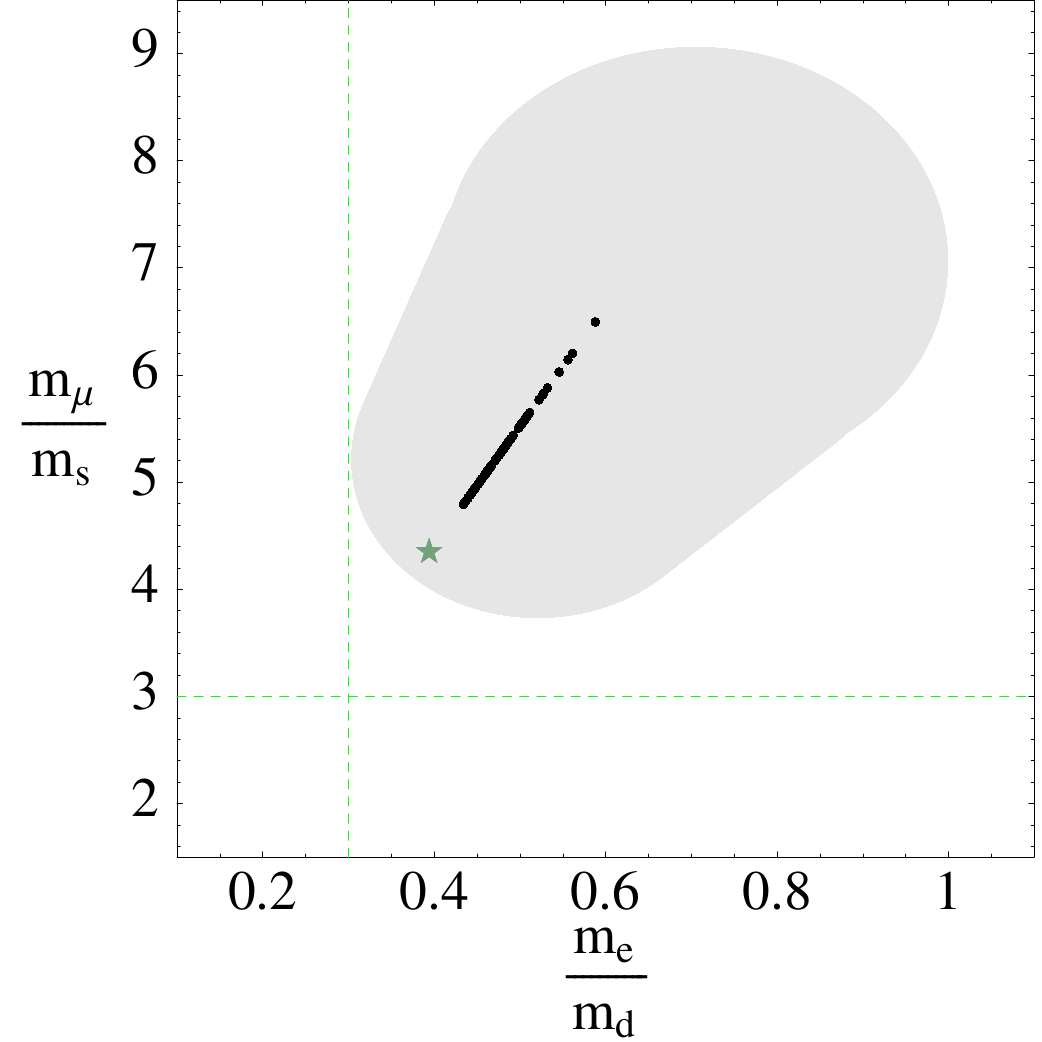}
   \includegraphics[scale=0.58]{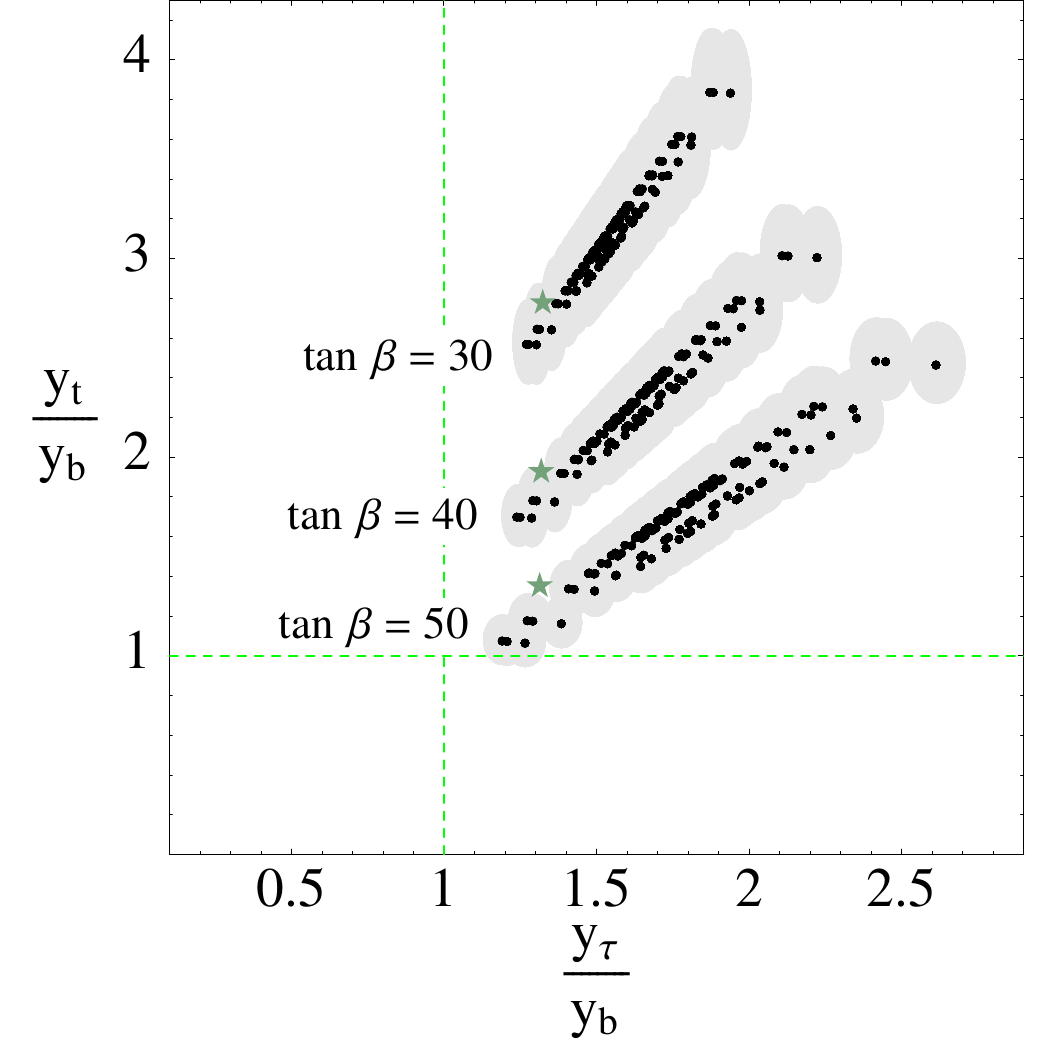}
   \includegraphics[scale=0.58]{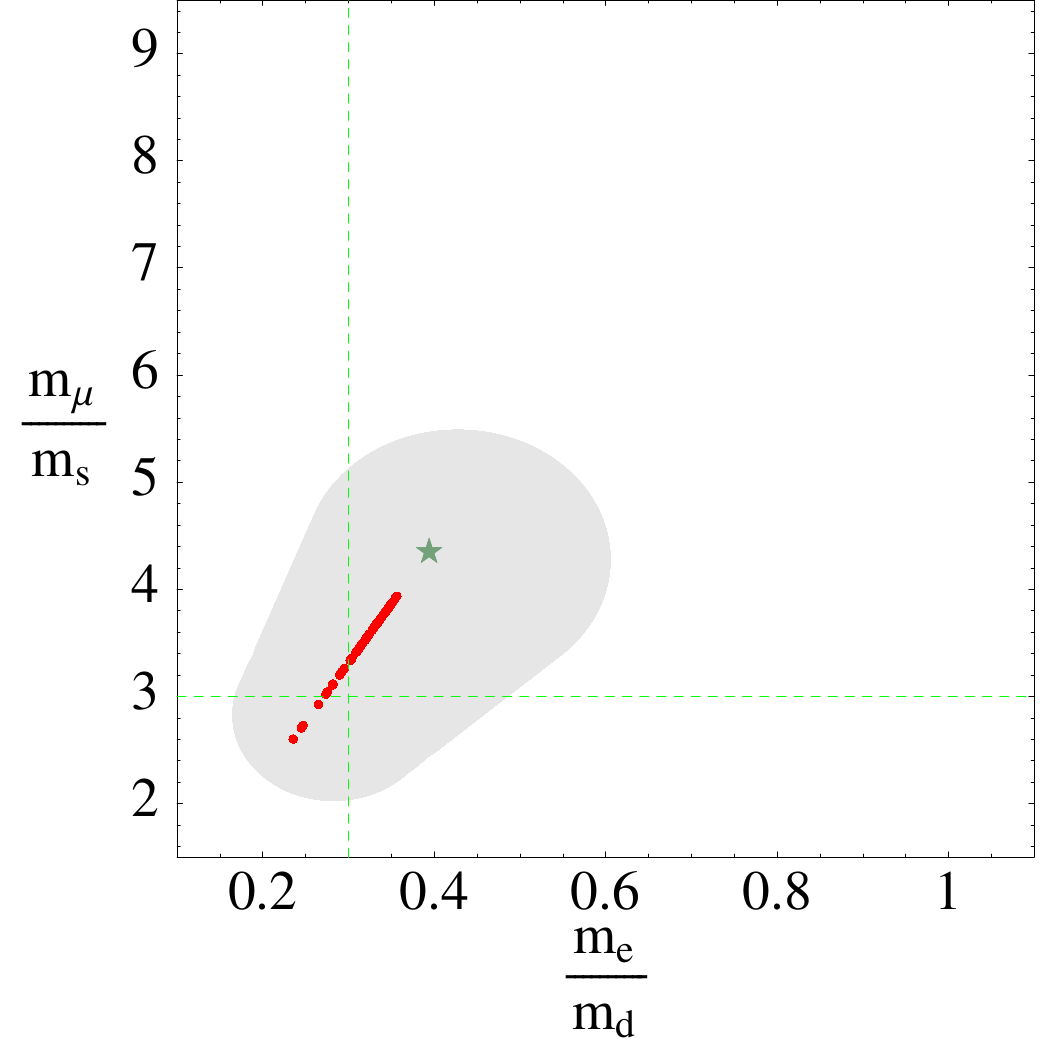}
   \includegraphics[scale=0.58]{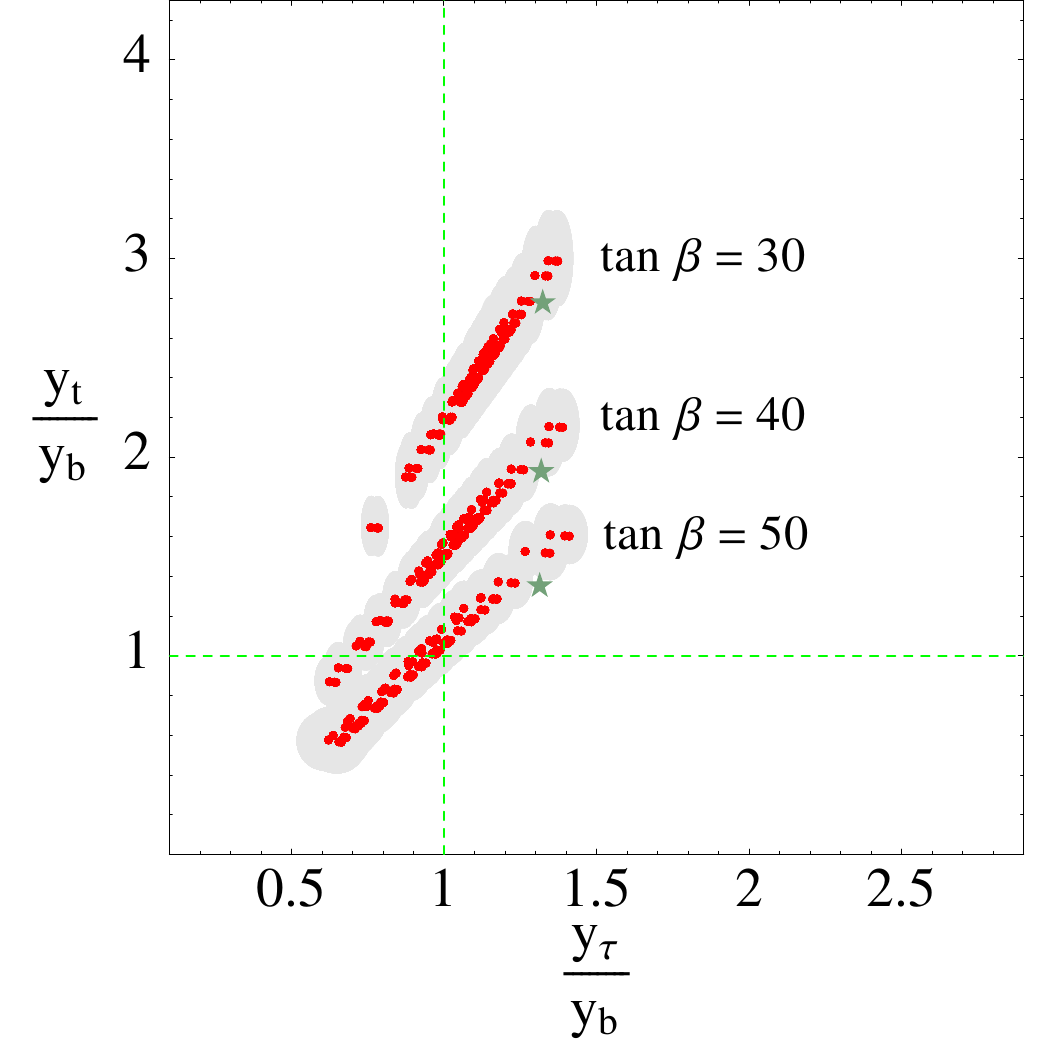}
   \includegraphics[scale=0.58]{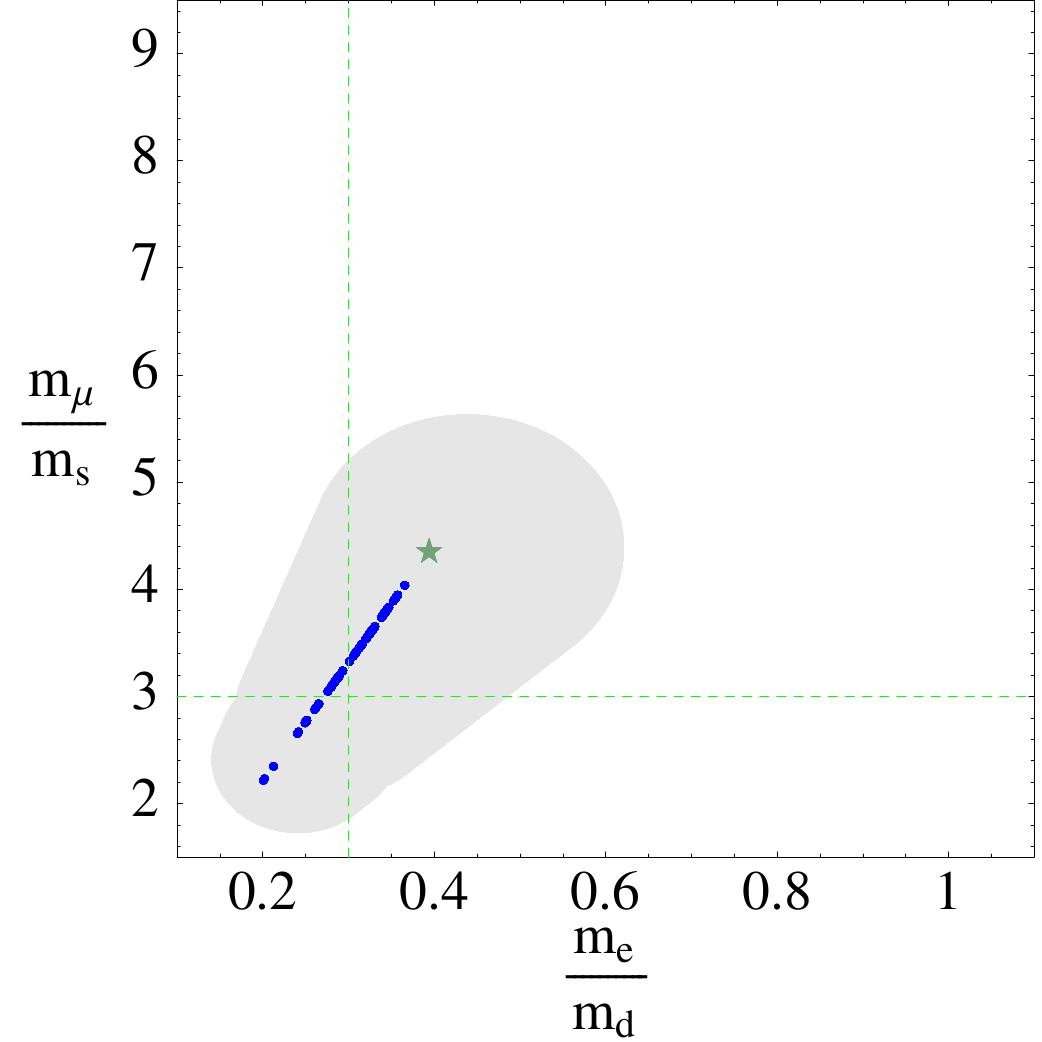}
   \includegraphics[scale=0.58]{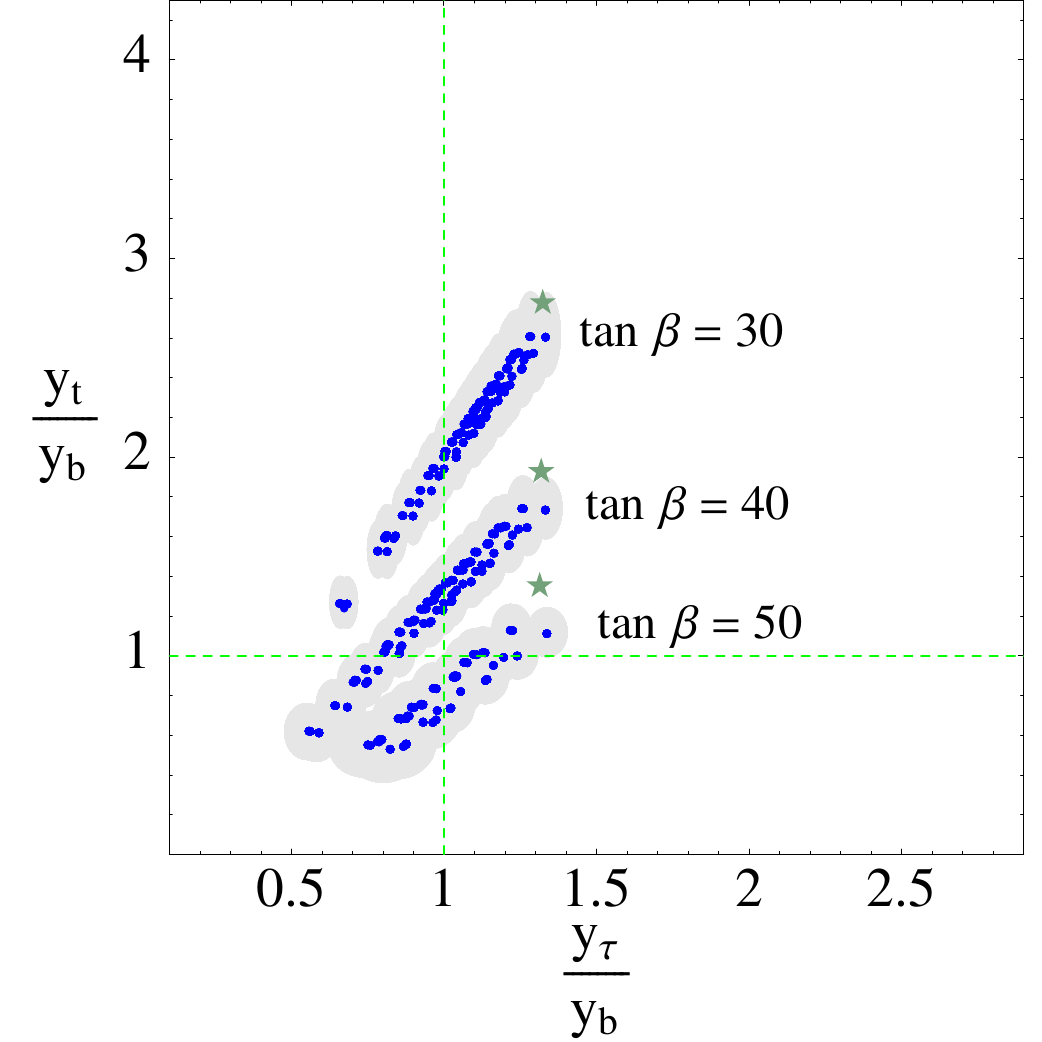}
   \caption{Scatter plots illustrating the ranges of GUT scale ratios $m_\mu/m_s$, $m_e/m_d$, $y_\tau/y_b$ and $y_t/y_b$ corresponding to the example ranges of SUSY parameters in table \ref{tab:SUSYpar} including $A_t = 0$ (first row: case $g_+$, second row: case $g_-$, third row: case $a$) for $\tan \beta = 30$, 40 and 50. The green dashed lines in the left plots correspond to the GJ relations. In the right plots they indicate the ratios $y_\tau/y_b=1$ and $y_t/y_b=1$. The green stars correspond to case 0 (no SUSY threshold corrections). The point-bands in the plots on the right correspond to the different values of $\tan \beta$. 
The grey areas correspond to the (1$\sigma$) quark mass errors for each shown data point.    \label{fig:Scatter}}   
\end{figure}

The results of our parameter scans are presented as scatter plots in figure~\ref{fig:Scatter}. 
The grey areas correspond to the (1$\sigma$) quark mass errors for each shown data point. 
In the first column, $m_s$ and $m_d$ have been varied, and  in the second column $y_b$ and $y_t$, using 
best-fit values for the remaining fermion masses. In comparison to the quark mass errors, the charged lepton mass errors are negligible. 
In figure~\ref{fig:Scatter} only those data points are shown where $y_b$, including its $1\sigma$ error, 
stays perturbative. 
For comparison we have also included the best-fit values, which would be obtained without SUSY threshold effects. It can be seen that, with SUSY threshold corrections included, the shown ratios are (for almost all parameter points) significantly shifted to large values for case $g_+$ and to smaller values for case $g_-$ and $a$.

The plots also reveal some interesting features, which we discuss now. For example we notice that small values for $m_\mu/m_s$ are correlated (nearly linearly) to small values of $m_e/m_d$. This correlation is connected to our assumption that the sfermion masses of the first two families are assumed to be identical. 
Because of the quark mass errors, this correlation is somewhat smeared.
Looking next at the third family relations $y_\tau/y_b$ versus $y_t/y_b$, we find that there is an additional $\tan \beta$ dependence, which is due to the fact that the relation of top mass and Yukawa coupling differs from the relations for down-type quarks and charged leptons by a factor of $\tan \beta$. For all three cases we can therefore distinguish three bands, which correspond to the three values of $\tan \beta$ in our parameter scan. The scans show that for case $g_-$ and $a$ it is in principle possible to obtain third family Yukawa unification for $\tan \beta \approx 50$ (in fact for $\tan \beta$ somewhat below $50$ in case $a$), whereas for $g_+$ we found that it could not be exactly realised. Although $y_b \approx y_t$ can be achieved for $\tan \beta = 50$, $A_t = -1$~TeV, $\mu = 0.5$~TeV, light gaugino masses, $m_{\tilde{d}_3} = 1.5$ TeV and $m_{\tilde{Q}_3} = m_{\tilde{u}_3} = 0.5$~TeV, we found $y_\tau/y_b \gtrsim 1.1$ in the considered parameter range. 

We note that since we have not specified the remaining SUSY parameters (which do not enter the formulae for the threshold corrections) we have not applied various relevant phenomenological constraints on the spectrum.
We would therefore like to warn the reader that for additional assumptions for the SUSY parameters some of the considered parameter points may be phenomenologically challenged already by the present experimental data (e.g.\ from $B$ physics, $g_{\mu} - 2$ or from LFV charged lepton decays in seesaw scenarios). In particular the cases with large $\tan \beta$, small pseudoscalar Higgs mass and large $|A_t|$ may (under some conditions) be challenged by $B_s \to \mu^+ \mu^-$ data and the case $\mu < 0$ (and $M_2 > 0$) may seem already disfavoured by $g_{\mu} - 2$ if the assumption is made that the ($\gtrsim 3 \sigma$) deviation from the SM prediction is restored by SUSY loop effects.

The ranges for the quark and lepton Yukawa couplings at the GUT scale and for the ratios of interest are presented in table \ref{tab:serror} without quark mass errors. They can be compared to the results of our semianalytical treatment  given in table \ref{tab:GJRana}. Comparing the two tables one can see that the ``mean values'' of the ranges agree well for the first two generations, however the extremal values are somewhat different. 
This is no surprise since in our naive estimates we have generically overestimated the ranges of the $\epsilon_i$ (since we have ignored possible correlations between the corrections). 
For the third generation, the ranges (boundaries) are slightly shifted. This effect is caused by the modified RG running with SUSY threshold corrections included. For the first two generations this effect is smaller due to the smaller Yukawa couplings. 

While the quark mass errors are still ignored in tables \ref{tab:GJRana} and \ref{tab:serror}, they are included in our final results listed in tables \ref{tab:Ratfinal} and \ref{tab:Yukfinal}. 
Here we have varied all the quark mass errors simultaneously. 
Comparing case 0 (without SUSY threshold corrections) to the cases $g_\pm$ and $a$, we see that the ranges for all types of Yukawa couplings, for down-type quarks, charged leptons and up-type quarks, as well as for all three generations, can be significantly affected by the SUSY threshold corrections. We note that for the up-type quarks, the changes are indirect in the sense that they are induced by modified RG running (mainly) due to the corrected $b$-quark and $\tau$-lepton Yukawa couplings.

\begin{table}
\begin{center}
\begin{tabular}{|c|c|c|c|c|}
\hline 				& Case 0 & Case $g_+$ & Case $g_-$ & Case $a$ \\ \hline
\hline $m_e/m_d$ 		& 0.39 & [0.45, 0.55] & [0.26, 0.34] & [0.24, 0.36] \\ 
\hline $m_\mu/m_s$ 		& 4.35 & [4.95, 6.03] & [2.92, 3.80] & [2.65, 3.92] \\ 
\hline $y_\tau/y_b$ 		& 1.32 & [1.23, 2.23] & $\leq 1.40$  & $\leq 1.35$  \\ 
\hline $y_t/y_b$ 		& 1.93 & [1.67, 3.03] & $\leq 2.17$  & $\leq 1.76$  \\ \hline
\hline $y_e$ in $10^{-4}$	& 0.88 & [0.87, 0.97] & [0.80, 1.09] & [0.92, 1.44] \\
\hline $y_\mu$ in $10^{-2}$	& 1.85 & [1.83, 2.06] & [1.68, 2.31] & [1.95, 3.05] \\
\hline $y_\tau$			& 0.34 & [0.34, 0.39] & [0.31, 0.44] & [0.36, 0.60] \\ \hline
\hline $y_d$ in $10^{-4}$	& 2.22 & [1.66, 2.10] & [2.39, 3.80] & [2.73, 5.73] \\
\hline $y_s$ in $10^{-2}$	& 0.43 & [0.32, 0.40] & [0.46, 0.73] & [0.52, 1.10] \\
\hline $y_b$			& 0.26 & [0.16, 0.30] & $\geq 0.23$  & $\geq 0.29$  \\ \hline
\hline $y_u$ in $10^{-6}$	& 2.75 & [2.73, 2.76] & [2.74, 2.83] & [2.75, 2.87] \\
\hline $y_c$ in $10^{-3}$	& 1.34 & [1.33, 1.35] & [1.34, 1.38] & [1.34, 1.40] \\
\hline $y_t$			& 0.50 & [0.48, 0.51] & [0.50, 0.58] & [0.50, 0.62] \\ \hline
\end{tabular}
\end{center}
\caption{Ranges for the GUT scale ratios and Yukawa couplings (corresponding to the example ranges of SUSY parameters in table \ref{tab:SUSYpar}, including additionally $A_t = 0$) from our numerical analysis with $\tan \beta = 40$. Case 0 refers to the case without SUSY threshold corrections. Quark mass errors are not yet included. The results can be compared with the semianalytic estimates of table \ref{tab:GJRana}. Our final results, which include the experimental mass errors, are given in tables \ref{tab:Ratfinal} and \ref{tab:Yukfinal}.
Where $y_b$ becomes nonperturbatively large we have given only the boundary, for which $y_b$ stays perturbative up to the GUT scale.
\label{tab:serror}}
\end{table}

\section{Implications for GUT Model Building}\label{Sec:models}

In the previous sections we have seen that comparatively wide ranges of Yukawa couplings are allowed at the GUT scale, if possible SUSY threshold corrections are taken into account. From todays perspective, since we do not have any experimental confirmation for low-energy SUSY and therefore no knowledge of the sparticle spectrum, this can have a large impact for GUT model building. 
For instance, as has been shown in \cite{Ross:2007az}, the threshold corrections can make the GJ relations consistent with the latest experimental data on quark masses. If the GJ relations are assumed at high energies, this can be understood as a constraint on the SUSY parameter space and points to scenarios with $M_3 < 0$ and $\mu > 0$ (if phenomenological consistency with experimental results on $(g-2)_\mu$ is assumed to be restored by SUSY loops \cite{gmu}).
Another interesting aspect is that the wide allowed ranges for the Yukawa couplings at the GUT scale open up new possibilities for constructing SUSY GUT models to address the flavour problem. 

One example for an application of such alternative GUT scale ratios $m_\mu/m_s$ and $m_e/m_d$ can be found in \cite{Antusch:2005ca}, where an approach has been presented to realise the phenomenologically successful  relation $\theta_{12}+\theta_C \approx \pi/4$ (so-called quark-lepton complementarity \cite{QLC}) in unified theories. In this approach, the Yukawa matrices for the charged leptons and down-type quarks emerge from the identical higher-dimensional operators where quarks and leptons are unified in representations of the Pati-Salam gauge group $G_{4221}$ = SU(4)$_\mathrm{C} \times$ SU(2)$_\mathrm{L} \times$ SU(2)$_\mathrm{R} \times$ U(1)$_{\mathrm{B-L}}$. After spontaneous breaking of $G_{4221}$ to the SM gauge group Clebsch-Gordan factors lead to different (GUT scale) values for the charged lepton and down-type quark Yukawa couplings. In the approach of \cite{Antusch:2005ca}, for example, $m_\mu/m_s = 2$ was postulated at $M_\mathrm{GUT}$. 

More generally, the assumption that the Yukawa matrices for the charged leptons and down-type quarks are generated from the same set of higher-dimensional operators in quark-lepton unified theories leads to a large variety of possible ratios $m_\mu/m_s$ and $m_e/m_d$ which correspond to different choices of operators and their associated Clebsch factors. A table with a collection of possible Clebsch factors in the context of Pati-Salam theories can be found in the appendix of \cite{Allanach:1996hz}. Any of these combinations of Clebsch factors which results in ratios $m_\mu/m_s$ and $m_e/m_d$ consistent with the ranges in table \ref{tab:Ratfinal} are {\em a priori} interesting new options for GUT model building. On the other hand, values of $m_\mu/m_s$ larger than $3$ seem to be difficult to realise within the Pati-Salam framework, which may point to models where the leading contributions to charged lepton and down-type quark Yukawa matrices emerge from different operators.  

Finally, if low-energy SUSY is found at the LHC and if the SUSY parameters are determined at the LHC (and/or ILC), the SUSY threshold corrections can be calculated and significantly more precise statements about the GUT scale values of the quark and lepton masses can be achieved. 
In addition, more precise values of the low-energy quark masses (in particular of the first two generations) would be highly desirable to improve the GUT scale predictions. 
This would allow to select among the possible GUT scale ratios of quark and lepton Yukawa couplings and may point to unexpected new relations.

\section{Summary and Conclusions}

\begin{table}
\begin{center}
\begin{tabular}{|c|c|c|c|c|c|}
\hline  			& Ratio 	& Case 0 & Case $g_+$ & Case $g_-$ & Case $a$ \\ \hline
\hline $\tan \beta = 30$ 	& $m_e/m_d$	& [0.28, 0.67] & [0.30, 0.86] & [0.21, 0.61] & [0.20, 0.62] \\
\cline{2-6} 			& $m_\mu/m_s$	& [3.39, 6.07] & [3.73, 7.79] & [2.54, 5.49] & [2.40, 5.63] \\
\cline{2-6} 			& $y_\tau/y_b$	& [1.27, 1.38] & [1.20, 2.02] & [0.71, 1.43] & [0.60, 1.39] \\
\cline{2-6} 			& $y_t/y_b$	& [2.56, 3.02] & [2.36, 4.19] & [1.50, 3.28] & [1.14, 2.87] \\ \hline
\hline $\tan \beta = 40$ 	& $m_e/m_d$	& [0.28, 0.67] & [0.31, 0.93] & [0.19, 0.59] & [0.17, 0.60] \\
\cline{2-6} 			& $m_\mu/m_s$	& [3.39, 6.07] & [3.85, 8.41] & [2.28, 5.30] & [2.07, 5.47] \\
\cline{2-6} 			& $y_\tau/y_b$	& [1.26, 1.38] & [1.16, 2.32] & $\leq 1.46$  & $\leq 1.41$  \\
\cline{2-6} 			& $y_t/y_b$	& [1.77, 2.11] & [1.55, 3.31] & $\leq 2.38$  & $\leq 1.94$  \\ \hline
\hline $\tan \beta = 50$ 	& $m_e/m_d$	& [0.28, 0.67] & [0.32, 1.00] & [0.16, 0.57] & [0.14, 0.59] \\
\cline{2-6} 			& $m_\mu/m_s$	& [3.39, 6.07] & [3.98, 9.06] & [2.02, 5.12] & [1.72, 5.31] \\
\cline{2-6} 			& $y_\tau/y_b$	& [1.25, 1.38] & [1.08, 2.73] & $\leq 1.49$  & $\leq 1.43$  \\
\cline{2-6} 			& $y_t/y_b$	& [1.22, 1.50] & [0.94, 2.74] & $\leq 1.81$  & $\leq 1.31$  \\ \hline
\end{tabular}
\end{center}
\caption{Ranges for the GUT scale ratios $m_\mu/m_s$, $m_e/m_d$, $y_\tau/y_b$ and $y_t/y_b$ corresponding to the example ranges of SUSY parameters $g_+$, $g_-$ and $a$ defined in table \ref{tab:SUSYpar} including $A_t = 0$. The results have been extracted from the numerical analysis (parameter scan) with $\tan \beta = 30,40$ and $50$, where in addition to the SUSY threshold corrections the present experimental errors for the quark masses have been included. Case 0 refers to the case without SUSY threshold corrections.
Where $y_b$ becomes nonperturbatively large we have given only the boundary, for which $y_b$ stays perturbative up to the GUT scale.
\label{tab:Ratfinal}}
\end{table}

\begin{table}
\begin{center}
\begin{tabular}{|c|c|c|c|c|c|}
\hline 				& Yukawa 		& Case 0       & Case $g_+$ & Case $g_-$ & Case $a$ \\ \hline
\hline $\tan \beta = 30$ 	& $y_e$ in $10^{-4}$	& 0.62         & [0.62, 0.67] & [0.58, 0.66] & [0.63, 0.79] \\
\cline{2-6} 			& $y_\mu$ in $10^{-2}$	& [1.30, 1.32] & [1.32, 1.41] & [1.22, 1.40] & [1.34, 1.66] \\
\cline{2-6} 			& $y_\tau$		& 0.23         & [0.23, 0.25] & [0.22, 0.25] & [0.24, 0.30] \\ \cline{2-6}
\cline{2-6} 			& $y_d$ in $10^{-4}$	& [0.92, 2.26] & [0.75, 2.14] & [0.98, 3.11] & [1.06, 3.89] \\
\cline{2-6} 			& $y_s$ in $10^{-2}$	& [0.21, 0.39] & [0.17, 0.37] & [0.23, 0.53] & [0.25, 0.67] \\
\cline{2-6} 			& $y_b$			& [0.17, 0.18] & [0.12, 0.20] & [0.16, 0.34] & [0.18, 0.48] \\ \cline{2-6}
\cline{2-6} 			& $y_u$ in $10^{-6}$	& [1.79, 3.88] & [1.78, 3.89] & [1.78, 3.93] & [1.79, 3.98] \\
\cline{2-6} 			& $y_c$ in $10^{-3}$	& [1.14, 1.54] & [1.13, 1.54] & [1.14, 1.56] & [1.14, 1.58] \\
\cline{2-6} 			& $y_t$			& [0.46, 0.51] & [0.46, 0.52] & [0.46, 0.54] & [0.46, 0.57] \\ \hline
\hline $\tan \beta = 40$ 	& $y_e$ in $10^{-4}$	& [0.87, 0.88] & [0.86, 0.99] & [0.79, 1.62] & [0.91, 1.77] \\
\cline{2-6} 			& $y_\mu$ in $10^{-2}$	& [1.83, 1.87] & [1.82, 2.08] & [1.67, 3.43] & [1.93, 3.74] \\
\cline{2-6} 			& $y_\tau$		& [0.34, 0.35] & [0.34, 0.39] & [0.30, 0.67] & [0.36, 0.76] \\ \cline{2-6}
\cline{2-6} 			& $y_d$ in $10^{-4}$	& [1.30, 3.21] & [0.97, 3.05] & [1.40, 8.41] & [1.59, 9.81] \\
\cline{2-6} 			& $y_s$ in $10^{-2}$	& [0.30, 0.55] & [0.23, 0.52] & [0.33, 1.44] & [0.37, 1.69] \\
\cline{2-6} 			& $y_b$			& [0.25, 0.27] & [0.15, 0.33] & $\geq 0.22$  & $\geq 0.27$  \\ \cline{2-6}
\cline{2-6} 			& $y_u$ in $10^{-6}$	& [1.79, 3.91] & [1.78, 3.92] & [1.79, 4.07] & [1.80, 4.15] \\
\cline{2-6} 			& $y_c$ in $10^{-3}$	& [1.14, 1.55] & [1.14, 1.56] & [1.14, 1.62] & [1.14, 1.65] \\
\cline{2-6} 			& $y_t$			& [0.47, 0.53] & [0.46, 0.54] & [0.47, 0.65] & [0.48, 0.72] \\ \hline
\hline $\tan \beta = 50$ 	& $y_e$ in $10^{-4}$	& [1.18, 1.23] & [1.13, 1.53] & [1.04, 2.31] & [1.30, 3.80] \\
\cline{2-6} 			& $y_\mu$ in $10^{-2}$	& [2.50, 2.60] & [2.39, 3.24] & [2.19, 4.89] & [2.74, 8.04] \\
\cline{2-6} 			& $y_\tau$		& [0.50, 0.52] & [0.47, 0.69] & [0.42, 1.07] & [0.56, 2.20] \\ \cline{2-6}
\cline{2-6} 			& $y_d$ in $10^{-4}$	& [1.77, 4.46] & [1.20, 4.56] & [1.91, 12.22]& [2.36, 22.68]\\
\cline{2-6} 			& $y_s$ in $10^{-2}$	& [0.41, 0.77] & [0.28, 0.78] & [0.44, 2.10] & [0.55, 3.90] \\
\cline{2-6} 			& $y_b$			& [0.36, 0.42] & [0.19, 0.60] & $\geq 0.30$  & $\geq 0.43$  \\ \cline{2-6}
\cline{2-6} 			& $y_u$ in $10^{-6}$	& [1.81, 3.95] & [1.79, 4.00] & [1.80, 4.18] & [1.81, 4.21] \\
\cline{2-6} 			& $y_c$ in $10^{-3}$	& [1.15, 1.57] & [1.14, 1.59] & [1.15, 1.66] & [1.16, 1.67] \\
\cline{2-6} 			& $y_t$			& [0.49, 0.56] & [0.46, 0.59] & [0.48, 0.78] & [0.50, 0.86] \\ \hline
\end{tabular}
\end{center}
\caption{Ranges for the GUT scale values of the Yukawa couplings corresponding to the example ranges of SUSY parameters $g_+$, $g_-$ and $a$ defined in table \ref{tab:SUSYpar} including $A_t = 0$. The results have been extracted from the numerical analysis (parameter scan) with $\tan \beta = 30,40$ and $50$, where in addition to the SUSY threshold corrections the present experimental errors for the quark masses have been included. Case 0 refers to the case without SUSY threshold corrections.
Where $y_b$ becomes nonperturbatively large we have given only the boundary, for which $y_b$ stays perturbative up to the GUT scale.
\label{tab:Yukfinal}}
\end{table}

In this study we have investigated the effect of SUSY threshold corrections on the values of the running quark and charged lepton Yukawa couplings at the GUT scale within the MSSM in the large $\tan \beta$ regime. We have therefore solved the set of RGEs from the top mass scale $m_{t}$ to the GUT scale $M_\mathrm{GUT}$, taking into account the one-loop threshold corrections at a SUSY scale ($M_\mathrm{SUSY}$) which has been assumed to be about $1$ TeV. With superparticle masses not too widely split around $M_\mathrm{SUSY}$, RG running can be performed in a two-step procedure, first with the SM RGEs from $m_t$ to $M_\mathrm{SUSY}$, where one-loop matching is performed, and then from $M_\mathrm{SUSY}$ to $M_\mathrm{GUT}$ using MSSM RGEs. The additional right-handed neutrino thresholds within the seesaw framework have also been considered but turned out to have no significant effects.  

At the matching scale $M_\mathrm{SUSY}$, we have included all relevant $\tan \beta$-enhanced one-loop corrections to the quark and lepton Yukawa couplings, which stem from SUSY QCD contributions (for all down-type quarks), SUSY EW contributions (for all down-type quarks and charged leptons) and from the trilinear soft SUSY breaking coupling $A_t$ (for the $b$ quark). For this purpose we have also calculated the SUSY EW corrections to the charged lepton Yukawa couplings in the EW unbroken phase. In general they cannot be neglected in a quantitative analysis of the SUSY threshold effects as we have shown.

Using the above described approach, we have analysed which values of quark and charged lepton masses may be realised at $M_\mathrm{GUT}$. Since the possible values depend on a large number of SUSY parameters as well as on the quark mass errors, we have organised our analysis as follows:

\begin{itemize}

\item We have first analysed the size of the SUSY threshold corrections for three example ranges of SUSY parameters $g_+$, $g_-$ and $a$ (listed in table \ref{tab:SUSYpar}). While there is a direct dependence on the gaugino masses and on $\mu$ in the corresponding contributions in equations (\ref{eq:expq}) to (\ref{eq:expl_end}), one can see that the dependence on the squark and slepton masses is somewhat hidden
in the function $H_2$. We have used the ranges for the threshold corrections to estimate the possible variety of GUT scale values of the Yukawa couplings due to the unknown SUSY spectrum semianalytically.

\item We have next used best-fit values for the low-energy quark and lepton masses and isolated the effects of $\tan \beta$ and $|\mu|$ as well as of the signs of $\mu$ and of the gluino mass $M_3$ (keeping $M_1$ and $M_2$ positive). 
On the one hand this illustrates the dependence on $\tan \beta$ and $|\mu|$ anticipated from the leading SUSY QCD contribution in equation (\ref{eq:expq}). 
On the other hand, comparing the cases $\mu > 0$, $M_3 < 0$ to $\mu < 0$, $M_3 > 0$ (with sparticle masses fixed at this stage at $1$ TeV) demonstrates the importance of the EW contributions to the threshold corrections. Our results are shown as contour plots in figures~\ref{fig:Contour12} and \ref{fig:Contour34}. 

\item We have then discussed the effects of $A_t$, which can provide an important contribution to the correction for the $b$ quark Yukawa coupling $y_b$ (but can be neglected for the others). 
Furthermore, varying $M_\mathrm{SUSY}$ (but keeping it still around the TeV scale) did in general not have a significant effect. Similarly, including right-handed neutrino thresholds, which appear in seesaw scenarios for small neutrinos masses, did not significantly affect the GUT scale values of the quark and charged lepton Yukawa couplings, at least compared to the relevant SUSY parameters.

\item Finally we have turned to the numerical discussion of the impact of the sparticle spectrum using the same example ranges of SUSY parameters  as for the semianalytic estimates.
We note that since we have not specified the remaining SUSY parameters (which do not enter the formulae for the threshold corrections) we have not applied various relevant phenomenological constraints on the spectrum (cf.~discussion in section \ref{Sec:Scan}). The numerical results for our example ranges of SUSY parameters $g_+$, $g_-$ and $a$ are summarised as scatter plots in figure \ref{fig:Scatter} for $\tan \beta = 30,$ 40 and 50 and additionally in table~\ref{tab:serror} for $\tan \beta = 40$.

\end{itemize}

Our final results, with quark errors included, are presented in tables~\ref{tab:Ratfinal} and \ref{tab:Yukfinal}. They provide the resulting possible ranges for quark and charged lepton Yukawa couplings at the GUT scale in the presence of threshold corrections (for the example SUSY parameter ranges $g_+$, $g_-$ and $a$) as well as the resulting ranges for several relevant quark and lepton mass ratios. The tables illustrate that with SUSY threshold effects (and quark mass errors) included, a wide range of GUT scale values of down-type quark and charged lepton Yukawa couplings could be realised, consistent with the low-energy experimental data on quark and charged lepton masses. One interesting aspect is that this opens up new possibilities for constructing GUT models of fermion masses and mixings. SUSY threshold corrections can on the one hand improve consistency with existing postulated GUT scale relations such as the GJ relations $m_\mu/m_s = 3$, $m_e/m_d = 1/3$ or third family Yukawa unification $y_t = y_b = y_\tau$, but they might also point to new GUT scale relations (cf.~discussion in section \ref{Sec:models}), if low-energy supersymmetry is discovered at the LHC.

\section*{Acknowledgements}
This work was partially supported by The Cluster of Excellence for
Fundamental Physics ``Origin and Structure of the Universe'' (Garching and 
Munich).

\providecommand{\bysame}{\leavevmode\hbox to3em{\hrulefill}\thinspace}

\end{document}